\documentclass[10pt,journal,compsoc]{IEEEtran}

\ifCLASSOPTIONcompsoc
  \usepackage[nocompress]{cite}
\else
  \usepackage{cite}
\fi

\usepackage{array}

\ifCLASSOPTIONcompsoc
  \usepackage[caption=false,font=footnotesize,labelfont=sf,textfont=sf]{subfig}
\else
  \usepackage[caption=false,font=footnotesize]{subfig}
\fi

\usepackage{inputenc}
\usepackage[english]{babel}
\usepackage[dvipsnames]{xcolor}
\usepackage{graphicx} 
\usepackage{caption}
\usepackage{amsmath, amsfonts}
\usepackage{booktabs}
\usepackage{diagbox}
\usepackage{algorithm}
  \newtheorem{teo}{Theorem}

  \newtheorem{defi}{Definition}

  \newtheorem{remark}{Remark}

 \newcommand*{\defeq}{\mathrel{\vcenter{\baselineskip0.5ex \lineskiplimit0pt
                     \hbox{\scriptsize.}\hbox{\scriptsize.}}}%
                     =}

  \makeatletter
\def\ps@pprintTitle{%
   \let\@oddhead\@empty
   \let\@evenhead\@empty
   \def\@oddfoot{\reset@font\hfil\thepage\hfil}
   \let\@evenfoot\@oddfoot
}
\makeatother

\usepackage{hyperref}
\hypersetup{
    colorlinks=true,
    linkcolor=blue,
    filecolor=magenta,      
    urlcolor=cyan,
    }

\begin{document}

\title{High Dimensional Mode Hunting Using Pettiest Components Analysis}

\author{Tianhao~Liu,~%\IEEEmembership{Member,~IEEE,}
        Daniel Andr\'es D\'{\i}az-Pach\'on,~%\IEEEmembership{Fellow,~OSA,}
        J. Sunil Rao,~%\IEEEmembership{Life~Fellow,~IEEE}% <-this % stops a space
        and Jean-Eudes Dazard~
\thanks{JSR was partially supported by NSF grant DMS-1915976 and NIH grants U54 MD010722 and UL1 TR000460. DADP was partially supported by grant AWD-005895 from the Walter Bradley Center for Natural and Artificial Intelligence. \textit(Corresponding author: Daniel Andr\'es D\'{\i}az-Pach\'on.)}%
\IEEEcompsocitemizethanks{\IEEEcompsocthanksitem T. Liu, D. A. D\'{\i}az-Pach\'on and J. S. Rao are with the Division of Biostatistics, University of Miami, Miami, Fl, 33136\protect\\
E-mail: txl646@miami.edu, Ddiaz3@miami.edu, JRao@miami.edu
\IEEEcompsocthanksitem J-E. Dazard was with the Center of Proteomics and Bioinformatics at Case Western Reserve University, Cleveland, OH, 44106\protect\\% <-this % stops an unwanted space
E-mail: jean-eudes.dazard@case.edu}

}

\IEEEtitleabstractindextext{%
\begin{abstract}
Principal components analysis has been used to reduce the dimensionality of datasets for a long time. In this paper, we will demonstrate that in mode detection the components of smallest variance, the pettiest components, are more important. We prove that for a multivariate normal or Laplace distribution, we obtain boxes of optimal volume by implementing ``pettiest component analysis'', in the sense that their volume is minimal over all possible boxes with the same number of dimensions and fixed probability. This reduction in volume produces an information gain that is measured using active information. We illustrate our results with a simulation and a search for modal patterns of digitized images of hand-written numbers using the famous MNIST database; in both cases pettiest components work better than their competitors. In fact, we show that modes obtained with pettiest components generate better written digits for MNIST than principal components.
\end{abstract}

}

\maketitle

\IEEEdisplaynontitleabstractindextext

\IEEEpeerreviewmaketitle

\IEEEraisesectionheading{\section{Introduction}\label{sec:introduction}}

\IEEEPARstart{P}{rincipal} components analysis is a widely used learning tool, particularly in unsupervised learning, to reduce the dimension of the space by projecting on the orthogonal rotation that maximizes the variance, especially in datasets where the number of variables is much larger than the number of observations \cite[Ch.~8]{Mardia1979}. In principal components regression, most practitioners prefer to discard the components of the input with the smallest variance, using just the first few principal components. Discarding the components of smallest variance does give satisfying results in some situations. In fact, Artemiu and Li gave conditions under which the first few leading principal components have a higher probability of correlating with the dependent variable than the small-variance ones \cite{ArtemiouLi2009}.  However, even though principal components regression can be useful, some caution is needed when it is implemented. 

As Jolliffe demonstrated in a classic paper, we can easily find some ordinary examples in which small-variance components become equally important, if not more, than large-variance ones \cite{Jolliffe1982}. Hadi and Ling have also presented three cautionary notes on using principal components regression, mainly due to issues arising from multicollinearity \cite{HadiLing1998}. In the end, the strongest reason to question principal components regression is that the dependent variables are never used in principal components analysis. In the words of Cox, it is hard to see any reason ``why the dependent variable should not be closely tied to the least important principal components'' \cite[p.~272]{Cox1968}. Here we will call those least important principal components \textit{pettiest components}.

Although these arguments are informative, to the best of our knowledge, few have tried to make a systematic use of the pettiest components in their own right. That is, most criticisms have focused on the negative aspect of finding isolated counter-examples showing that principal components regression is not desirable, based on the underlying assumption that the leading principal components are better, but the positive aspect of systematically selecting the pettiest components has received little attention. The exception seems to be a recent study by Sando and Hino where the authors propose a modal principal component analysis whose goal is to develop a robust PCA method. Through a kernel function they estimate the mode and use this estimation in order to find the pettiest components, that they call them ``minor components'' \cite{SandoHino2020}.

Our reasoning goes the other way around ---we take advantage from these pettiest components in order to find modes. We will show that the smallest regions of the same probability are achieved using the pettiest components (Section \ref{Sec:PetTheorems}). Intuitively, the result presented here follows this idea: In a space $S \subset \mathbb R^p$, define a $\beta$\textbf{-mode} of a continuous distribution as the region of the space with the smallest volume constrained to have a probability $\beta$. Assume the underlying distribution of a $p$-dimensional vector $\mathbf x$ is a multivariate normal, and that we project it to a $p'$-dimensional space, with $p' < p$. Then the box that minimizes the size, among all possible boxes of probability $\beta$, is the one projected over the subspace of the orthogonal variables with the smallest variance.

This work can be seen as a continuation of \cite{DiazDazardRao2017}, where the optimality, in terms of smallest volume, of the $p$-dimensional box centered around the mean in the direction of the eigenvectors was proved for a multivariate normal distribution. The next obvious step, the reduction of dimensionality to $p'$ dimensions, is accomplished here; but contrary to common practice, we show that the optimal $p'$-box is obtained when the $p$-dimensional box in the previous step is projected to the subspace of the $p'$ pettiest components. 

Why the interest in the $\beta$-regions with the smallest volume? An answer lies in the fact that such regions might contain a large amount of relative Shannon information among all the regions with probability $\beta$ (see Subsection \ref{Optimality}). Another answer is that, in spite of the $p$-dimensional support of a continuous distribution having uncountably infinite regions of probability $\beta$, even when $p=1$, the smallest region of probability $\beta$ will tell us that the data is highly concentrated inside such region. 

There is a vast number of applications where this is of interest. For instance, in more accurate predictions based on closeness to a given target, as in the Netflix recommendations algorithm; or in medical image recognition, where a high concentration of points around a small region can help with tumor detection. And principal components have been widely used for data compression and image reconstruction \cite[pp.~536-539]{HastieTibshiraniFriedman2009}, \cite[pp.~433-445]{VanderPlas2016}. This paper will exploit the link between modes, principal components analysis and image reconstruction, illustrating with the MNIST dataset that pettiest components do a better job than principal components both in finding modal patterns and in reconstructing the image through such patterns (Section \ref{Sec:MNIST}).

Yet, mode detection continues being a difficult problem, especially in multivariate settings. One of the most famous ways to approach it, specially in computer vision, is via the mean-shift algorithm by kernel density estimation \cite{FukunagaHostetler1975}. It works well when $p=2$ having amenable asymptotic properties \cite{Parzen1962, Valpine2004}. However, it becomes very slow when $p > 2$, though  the speed improves when the shape of the density is known \cite{CuleEtAl2010}. More recently, Ruzankin and Logashov introduced a fast mode estimator with time complexity $O(pn)$, whereas the time complexity of other estimators is not less than $O(pn^2)$, where $n$ is the number of observations \cite{RuzankinLogashov2020}.

The article is organized as follows. Section \ref{Sec:Basic} will introduce principal and pettiest components, setting the notation for the rest of the article; active information will be defined and used to obtain optimality of $\beta$-modes; finally, we will introduce the two algorithms that are going to be used throughout the paper: the Patient Rule Induction Method (PRIM) and a modification called fastPRIM. Section \ref{Sec:PetTheorems} will prove that the $\beta$-mode in the direction of pettiest components is optimal. Section \ref{Sec:Sim} presents a simulation using a multivariate normal distribution. Section \ref{Sec:MNIST} will illustrate our findings with the MNIST dataset. Section \ref{Sec:Disc} ends the paper with a discussion and some open problems for future research. 

\section{Basic notions}\label{Sec:Basic}

\subsection{Components}

Suppose there is a $p$-dimensional dataset of $n$ observations $\{ x_1, \ldots, x_p \}^n_1$ with covariance matrix $\Sigma$. The problem is to find a vector $\mathbf a$ maximizing $\mathbf a^T \Sigma \mathbf a$, subject to $\mathbf a^T \mathbf a = 1$. By Lagrange multipliers, the original optimization problem is equivalent to solve the eigenvalue question $\Sigma \mathbf a = \lambda \mathbf a$. From the solution we get the eigenvectors $\mathbf a_1, \ldots, \mathbf a_p$, %sometimes called loadings, 
and respective eigenvalues $\lambda_1, \ldots, \lambda_p$. The original $p$-dimensional input space will be denoted by $\mathcal{X} (p)$, and the space after the rotation in the direction of its $p$ eigenvectors, will be denoted by $\mathcal{X}' (p)$.

We will call these eigenvectors \textit{components}, and their corresponding eigenvalue will be their variance. Then, sorting the components in order of increasing variance, the $k$ \textit{principal components} will be the $k$ components with the largest variances; and the $k$ \textit{pettiest components} will be the $k$ components with the smallest variance. 

\subsection{Optimality of the $\beta$-mode}\label{Optimality}

Several comments are in order regarding $\beta$-modes. First, finding a general solution for the set of minimum volume in the class $\mathcal C$ of all the Borel sets with probability $\beta$ in $\mathbb R^d$ seems in general intractable; therefore we consider the next best thing: the more manageable class $\mathcal C' \subset \mathcal C$ of all hyper-rectangles $I_1 \times \cdots I_p$, where $I_i$ are intervals in $\mathbb R$. Second, we talk about $\beta$-modes in general, instead of simply modes, because we are interested on regions of positive probability. Third, out of the previous considerations, there are no spaces without $\beta$-modes; i.e., even if $S$ is of finite volume and the underlying distribution is uniform we have uncountable $\beta$-modes, though these will not be very informative; on the other hand, if $S$ has infinite Lebesgue measure, a continuous distribution in $S$ will have at least a $\beta$-mode. And fourth, we talk about \textit a $\beta$-mode and not \textit{the} $\beta$-mode, because there can be more than one Borel set with identical hyper-volume and probability $\beta$ in the space; in fact, in many situations not only global but local $\beta$-modes are of interest. However, in the special case of unimodal distributions, with some care, it is possible to talk about {\it the} $\beta$-mode:

\begin{defi}[$\beta$-unimodality]
	We say that a continuous distribution $\phi$ is $\beta$-unimodal with $\beta$-mode $B$ if for any other $\beta$-mode $B'$ the volume of the symmetric difference $\text{Vol}(B \triangle B')$ is 0. In this case either $B$ or $B'$ can be taken as the $\beta$-mode.
\end{defi}

We now introduce active information to prove the optimality of the $\beta$-mode in terms of Shannon information. Active information represents the information change induced by $\phi$ with respect to the baseline distribution $\psi$  \cite{DiazSaenzRao2020}. Active information was introduced in the context of search problems and has also been used to detect modes in multivariate analyses \cite{DiazEtAl2019, DiazMarks2020a}.

\begin{defi}[Active information]
Let $\phi$ and $\psi$ be two continuous distributions on $\mathcal X(p)$. The active information of $\phi$ with respect to $\psi$ for an event $R \subset\mathcal X(p)$ is defined as
\begin{align}\label{actinfodef}
	\mathbf I_+(R) \defeq \log\frac{\phi(R)}{\psi (R)},
\end{align}
provided $R$ has positive probability under $\phi$ and $\psi$.
\end{defi}

\begin{teo}[Optimality of the $\beta$-mode]\label{optim}
	Let $\mathcal X(p)$ be bounded. Let $\phi$ be a continuous distribution on $\mathcal X(p)$. Then $B\subset \mathcal X(p)$ is the unique $\beta$-mode of $\phi$ if and only if $B$ maximizes the active information of $\phi$ with respect to a uniform distribution $\mathbf U$ among all the events of $\mathcal X(p)$ with probability $\beta$.
\end{teo}

\begin{IEEEproof}
	Take $B$ and $B'$ two events such that $\phi(B) = \phi(B') = \beta$, with respective volumes $v$ and $v'$.  Then the active information of $\phi$ with respect to $\mathbf U$ for the event $B$ is
	\begin{align}\label{actinfo}
		\mathbf I_+(B) \defeq \log\frac{\phi(B)}{\mathbf U (B)} = \log \frac{\beta}{v},
	\end{align}
	and the active information of $\phi$ with respect to $\mathbf U$ for the event $B'$ is
	\begin{align}\label{actinfo2}
		\mathbf I_+(B') \defeq \log\frac{\phi(B')}{\mathbf U (B')} = \log \frac{\beta}{v'}.
	\end{align}
	Therefore
	\begin{align}\label{actinfoComp}
		\mathbf I_+(B) - \mathbf I_+(B') &= \log \frac{\beta}{v} - \log \frac{\beta}{v'} \nonumber \\
								&=\log\frac{v'}{v},
	\end{align}
	which is positive if and only if $v' > v$.
\end{IEEEproof}

\subsection{Patient rule induction method}

The \textit{patient rule induction method} (PRIM) is a greedy algorithm designed for bump hunting \cite{FriedmanFisher1999}. A bump can be roughly defined as a region on the response variable with higher mean, compared to other places.

Assume we have a dataset $\{y, x \}^n_1$ where $x = \{ x_1, x_2, \ldots, x_p \}$ is a continuous random vector. $f(x)$ is the target function. The input domain $\mathcal X(p)$ is here called $S$. Defining $\bar f_B$ and $\bar f$ as
\begin{align*}
  \bar f_B &= \frac{\int_B f (x) p(x) dx}{\int_B p(x) dx}, \nonumber \\
  \bar f &= \int_S f (x) p (x) dx,
\end{align*}
where $p(x)$ is the density function of $x$, the method aims to find a box $B \subset S$ in which $\bar{f}_B / \bar{f}$ is maximized under the constraint that $B$ is not too small. In fact, the probability of $B$ is defined at the outset by a tuning meta-parameter $\beta$ for the smallest permitted box size (else one would always simply choose a zero measure set, making the numerator infinite without actually providing a useful answer). In practice, $\bar{f}_B / \bar{f}$ will be replaced by its corresponding estimator.

\begin{figure}[!t]
\centering 
\includegraphics[width=6.4cm,height=4.9	cm]{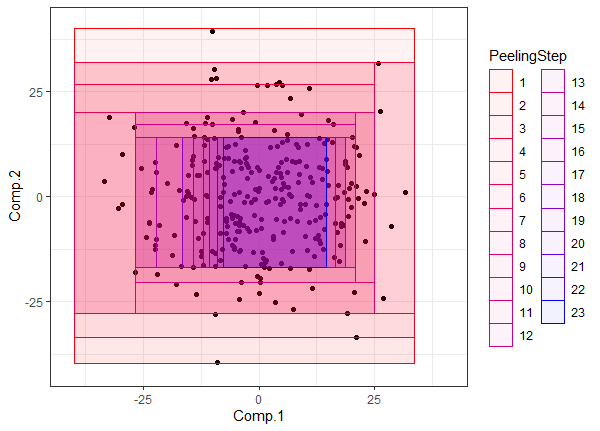}
\caption{The peeling procedure in PRIM.}
\label{Fig.peel}
\end{figure}

The algorithm is divided in three stages. First there is a \textit{peeling}: Beginning with the whole input space $S$, a whole class of eligible boxes for removal is set up. This class is made of sets of the form
\begin{align*}
 b_{j -} &= \{ x \mid x_j < x_{j (\alpha)} \},\\
  b_{j +} &= \{ x \mid x_j > x_{j (1 - \alpha)} \},
\end{align*}
where $x_{j (\alpha)}$ is an $\alpha$-quantile. If a box $b^\ast$ is in the eligible set, and without it the remaining $S \setminus b^\ast$ will give the maximum average value of response variable, $b^\ast$ is the specific box selected to remove in this loop. Formally we write
\begin{align*}
   b^{\ast} = \arg \max_b \text{avg} [y_i \mid x_i \in B - b],
\end{align*}
where avg is the average. Update the new box as $B = S \setminus b^{\ast}$ and run the same procedure on $B$. The process is iterated until the final box $B$ has probability $\beta$. The peeling procedure is shown in Fig. \ref{Fig.peel}. 

After peeling, a second step called \textit{pasting} is performed in order to correct for the greediness of peeling. We will not consider pasting here. Finally, the third step is called \textit{covering}, which means that after the peeling stage, the final box $B$ is deleted from $S$,  and then the whole procedure of peeling is repeated in $S \setminus B$.

PRIM has at least three shortcomings. First, it is computationally expensive; second, it does not behave well in the presence of collinearity; and third, even in a small number of dimensions, it cannot detect multiple bumps \cite{PolonikWang2010}. For this reason, \cite{DazardRao2010} developed a modified algorithm called \textit{local sparse bump hunting} (LSBH) in which the patient rule induction method is subsumed. The LSBH algorithm uses a recursive partition algorithm, say classification and regression trees, in order to divide the space $\mathcal X(p)$ in several regions $S_1, \ldots, S_R$; then applies sparse principal components analysis inside each particular region $S_i$ in order to reduce the dimension of each region in the partition; and finally applies the patient rule induction method to each rotated and projected subspace induced by $S_i$.  \cite{DazardRaoMarkowitz2012} applied LSBH to detect heterogeneity of colon tumors, dividing colon cancer patients into two subpopulations with different genetic/molecular profiles in all stages of cancer.

\subsection{FastPRIM}

Elaborating on LSBH, a new modified algorithm called \textit{fastPRIM} was developed in order to find the minimal volume of boxes with probability $\beta$ when the distribution is a multivariate normal \cite[Algorithm~4]{DiazDazardRao2017}. More accurately, the modes can be defined as $\beta$-modes; i.e., contiguous regions, not necessarily unique, with the smallest volume and probability $\beta$. Since the mode, the mean and the median of the normal distribution coincide, mode hunting in this case is equivalent to finding a box of probability $\beta$ centered around the mean. Explicitly, for $\alpha$ such that $(1-\alpha)^L = \beta$ and $B_0 = \emptyset$, fastPRIM works modifying PRIM as follows: 
\begin{enumerate}
	\item (Rotate) Generate $S_1 = \mathcal{X}' (p)$ (that is, rotate the space in the direction of its eigenvectors);
	\item For $j$ from 1 to $t$, 
		\begin{enumerate}
			\item (Peeling)  For $i$ from 1 to $L$, 
			
				Make $2p$ peelings from $S_j$ corresponding to each side of the box, each peeling having probability $\alpha(2p)^{-1}$. 
				
				Call $B_j$ the final box obtained after the 	$L$ peels.
			\item Set $S_{j+1} = S_j \setminus \bigcup_{k=0}^{j-1} B_k$.
		\end{enumerate}
	\item (Covering) Take $B = \bigcup_{k=1}^t B_k$. 
\end{enumerate}
Notice that after the $L$ stages of peeling, each $B_j$ is centralized around the mean and will have probability $\beta$. Thus the whole algorithm can be reduced to a single step. That is, the final box $B$ is centralized around the mean, each of its sides is parallel to an axis of the rotated space, and its vertices are located at the quantiles $2^{-1} (1 \pm \beta_T^{1/p})$ of the corresponding variable, where $\beta_T = \sum^t_{k = 1} \beta (1 - \beta)^{k - 1} = 1 - (1 - \beta)^t$ is the probability measure after $t$ steps of covering. As a result, since all the $2d$ sides are peeled with identical probability, $B$ is a rectangular Lebesgue set, it is centralized around the mean, and it is a square in probability (with each marginal having probability $\beta_T^{1/p}$).

\section{Pettiest components analysis with fastPRIM}\label{Sec:PetTheorems}

\cite[Proposition 1]{DiazDazardRao2017} proved that fastPRIM satisfies the following optimality property: when the distribution is a multivariate normal, say $\mathcal N(\mathbf 0, \Sigma)$, the box with the smallest volume subject to have probability $\beta$ is found in the direction of the rotation of the $p$ principal components; that is, the box centered around the mean of probability $\beta$ in $\mathcal X'(p)$. However, they did not prove any result dealing with dimension reduction. This article goes one step further: it shows that if we are going to consider reduction of dimensionality, say to $p' < p$, we should consider pettiest components instead of principal components.

Starting thus with the space $\mathcal X(p)$, let $B_i$ be the final box obtained by fastPRIM on an input space $\mathcal{X}'_i(p')$, where $i$ indicates a specific way of choosing $p'$ variables from $p$ variables. The collection of all such boxes will be $\mathcal B$. The Lebesgue measure of the box $B \in \mathcal B$ with probability measure $\beta$ is $\mathrm{Vol} (B | \beta)$. In the input space $\mathcal{X}'_i (p')$ spanned by the pettiest components, we write that specific box as $\mathbf B$. 

\begin{teo}\label{lema}
Let $X$ be a $p$-random vector such that its components $X_1, \ldots, X_p$ are independent of each other and have normal distribution $\mathcal{N} (0, \sigma_1^2), \ldots, \mathcal{N} (0,\sigma_p^2)$, respectively. Apply fastPRIM on each possible projection space $\mathcal X_i' (p')$ with same probability $\beta$. Then,
  \begin{align*}
  	\arg\min_{B \in \mathcal{B}} \mathrm{Vol} (B | \beta) = \mathbf B.
  \end{align*}
\end{teo}

\begin{IEEEproof}.
There are $\binom{p}{p'}$ ways to choose $p'$ dimensions from $p$. For $X_1, \ldots, X_p$ following independent normal distributions, we write $X_j \sim \mathcal N_j (0,\sigma_j^2)$, ($j=1, \ldots, p$). From the fact that the final box $B$ in fastPRIM is a square in probability, the marginal probability measure of every $B \in \mathcal B$ is $\beta^{1 / p'}$. Because all boxes $B \in \mathcal B$ are centralized at zero, the Lebesgue measure of $X_j$, subject to a probability $\beta^{1 / p'}$, can be calculated by the equation $P \{ - k \sigma_j < X_j < k \sigma_j \} = \beta^{1 / p'}$.  Here $k$ only depends on $\beta^{1 / p'}$ and it has nothing to do with $j$. Therefore, the edge length of $B_i$ in the $X_j$ direction is $2 k \sigma_j$. Then, 
\begin{align*}
	\mathrm{Vol} (B_i | \beta) = \prod^{p'}_{j^{(i)} = 1} 2 k\sigma_{j^{(i)}} = (2 k)^{p'} \prod^{p'}_{j^{(i)} = 1} \sigma_{j^{(i)}},
\end{align*} 
where $j^{(i)}$ denotes the $p'$ variables in the choice of $B_i$. Since a fixed $\beta$ implies a fixed $k$, the minimum of $\mathrm{Vol} (B_i | \beta)$ can thus be obtained by minimizing $\prod^{p'}_{j^{(i)} = 1} \sigma_{j^{(i)}}$. But this only requires the $\sigma_{1^{(i)}}, \ldots, \sigma_{p'^{(i)}}$ to have the smallest values, which is achieved by choosing the $X_{1^{(i)}}, \ldots, X_{p'^{(i)}}$ with the smallest variances. That is, the $p'$ pettiest components. The corresponding box found by fastPRIM with pettiest components is $\mathbf B$. 
\end{IEEEproof}

The basic idea behind the proof is to reduce the box to the marginals with same probability measure and then to prove a smaller variance corresponding to a smaller Lebesgue measure. Using Theorem \ref{lema}, we can easily show that the same conclusion is true for any multivariate normal distribution input under fastPRIM.

\begin{teo}\label{gentheo}
Let $X$ be a $p$-random vector distributed as $\mathcal{N}_p (0, \Sigma)$. Apply fastPRIM on each possible projection space $\mathcal X_i' (p')$ with same probability $\beta$. Then,
  \begin{align*}
  	\arg\min_{B \in \mathcal{B}} \mathrm{Vol} (B | \beta) = \mathbf B
  \end{align*}
 \end{teo}

\begin{IEEEproof}
For a multivariate normal distribution input, to rotate the space in the direction of its $p$ components is equivalent to solve the eigenvalue equation of covariance matrix $\Sigma$. The rotation will produce a diagonal matrix $D$ with eigenvalues as its diagonal elements. This new matrix $D$ is the covariance matrix of the components, which implies that they are all independent of each other and follow a normal distribution. So the problem is reduced to an instantiation of Theorem \ref{lema}. Thus the result holds as before.
\end{IEEEproof}

The multivariate normal distribution and its properties suggest that a more general result is attainable. It seems clear that when we have a symmetric multivariate distribution which is unimodal, and all its marginals belong to the same family of distributions, differing maybe only on the particular values of its parameters, a similar result to Theorem \ref{gentheo} can be obtained. For instance, Theorem \ref{laplace} below shows that a similar result is obtained for a symmetric multivariate Laplace distribution. The Laplace distribution is interesting in that even in the presence of 0 correlation of the marginals, which are univariate Laplace distributions themselves, they are not independent \cite[pp.~229-245]{KotzEtAl2001}. This fact highlights an important property about our results: they do not depend on the independence of the marginals but on their zero correlation, since we only need the variables being orthogonal.

\begin{teo}\label{laplace}
Let $X$ be a $p$-random vector with symmetric multivariate Laplace distribution. Assume that its components $X_1, \ldots, X_p$, being Laplace distributed, have parameters $\mathcal{L} (0, b_1), \ldots, \mathcal{L} (0, b_p)$, respectively. Assume also that $\text{corr}(X_i,X_j) = 0$ for all $i \neq j$, and $i,j \in \{1, \ldots, p\}$. Apply fastPRIM on each possible projection space $\mathcal X_i' (p')$ with same probability $\beta$. Then,
\begin{align*}
\arg\min_{B \in \mathcal{B}} \mathrm{Vol} (B | \beta) = \mathbf B.
\end{align*}
\end{teo}

\begin{IEEEproof}
The proof process is the analogous to the one in Theorem \ref{lema}, except that we apply it now to the Laplace distribution context. The goal is again to prove that smaller variance components will give smaller Lebesgue measures under the same probability measure $\beta^{1 / p'}$. The Lebesgue measure of the $j$-component can be calculated by $L_j=[Q (0, b_j ; 0.5 + \beta^{1 / p'}) - Q (0, b_j ; 0.5 - \beta^{1 / p'})]$, where $Q(\mu , b ;p) = \mu + b \ln (2p)$ for $p \leq 1/2$ and $\mu - b \ln (2-2p)$ for $p > 1/2$. Observe how variances influence $L_j$: a smaller variance $2b_j^2$ demands a smaller $b_j$. Since the quantile function $Q$ increases for $p \leq 1/2$ and decreases for $p > 1/2$, so $Q (0, b_j ; 0.5 + \beta^{1 / p'})$ will decrease and $Q (0, b_j ; 0.5 - \beta^{1 / p'})$ will increase, which makes $L_j$ decrease as a whole.
\end{IEEEproof}

\begin{remark}
Theorems \ref{gentheo} and \ref{laplace} suggest that our theorem can be further generalized to a broad range of multivariate distributions. In fact, the MNIST example considered below also points in the same direction, since the distributions are somewhat deviating from normality. However, it is important to notice that distributions pertaining to different families cannot in general be mixed and the previous results maintained. Consider, for instance, $X_1$ having normal distribution $\mathcal{N} (0, 1)$ and $X_2$ having Laplace distribution $\mathcal{L} (0, 1)$ (so its variance is 2), and the two variables are independent of each other. If these two represent the marginals of some multivariate distribution and we attempt to apply our result, we will obtain  that $L_1$ is going to be larger than $L_2$ when the probability measure is chosen to be small. Fig. \ref{counter} below illustrates clearly this point. However, it is also clear from Fig. \ref{counter} that there is a region  $R \subset [0,1]$ (not necessarily contiguous) such that for $\beta\in R$ it is better to select the normal distribution than the Laplace one.
\end{remark}

\begin{remark}
	Using Theorem \ref{optim}, Theorems  \ref{lema}, \ref{gentheo}, and \ref{laplace} show that the size of the box $\mathbf B$ maximizes the Shannon information among all the regions of probability $\beta$ in all projections $\mathcal X_i(p')$.
\end{remark}

\begin{remark}
The results obtained are distributional. However, as a corollary, from a sampling viewpoint fastPRIM is consistent: calculating the average of all the points inside the final centered box and taking this average as the center of our box, we know by the law of large numbers that the final box will be centered around the origin. Since the mean and the mode coincide in the normal distribution, as $n \rightarrow \infty$, the procedure is approaching a box whose center is the true mode. This is so with or without dimension reduction.
\end{remark}

\begin{figure}[!t]
\centering 
\subfloat[Distribution function]{\includegraphics[width=4cm,height=4.9cm]{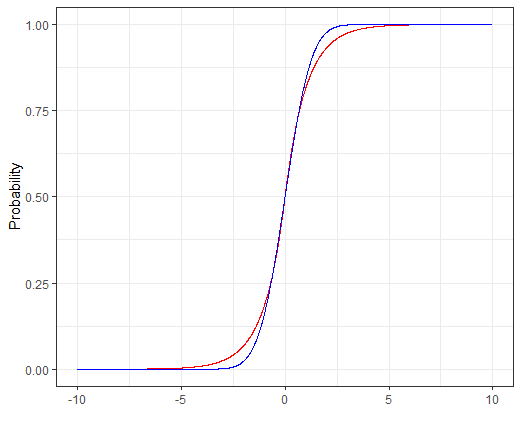}}
\subfloat[Zoom of distribution function]{\includegraphics[width=4cm,height=4.9cm]{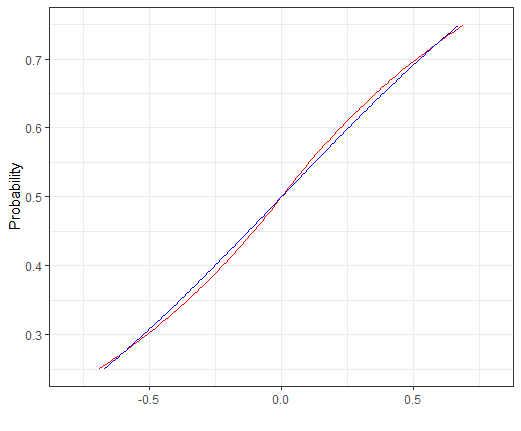}}\\
\caption{Counterexample for a mixed multivariate distribution. The blue line is normal $\mathcal{N}(0, 1)$ and the red line is Laplace $\mathcal{L} (0, 1)$.}\label{counter}
\end{figure}

\begin{remark}
The existence of the eigenvectors and the ordering of the eigenvalues are warranted by the conditions of Theorems \ref{gentheo} and \ref{laplace}. Thus, theoretically, components of zero variance are not under consideration by the hypotheses of normal and double exponential distributions. 
\end{remark}

\begin{remark}
 The number of pettiest components to be selected is an important question (as important as in principal components analysis, where it is still open). In applications, the threshold will likely require the input of the practitioner. Nonetheless, if there is geometric multiplicity for some eigenvalue producing that some vectors with the same variance are going to be left out whereas others are going to be selected, $p'$ can be modified either to include all the eigenvectors corresponding to such eigenvalue or to exclude them altogether. 
\end{remark}

\section{Simulation}\label{Sec:Sim}

\begin{figure}[!t]
\centering 
\includegraphics[width=4cm,height=3cm]{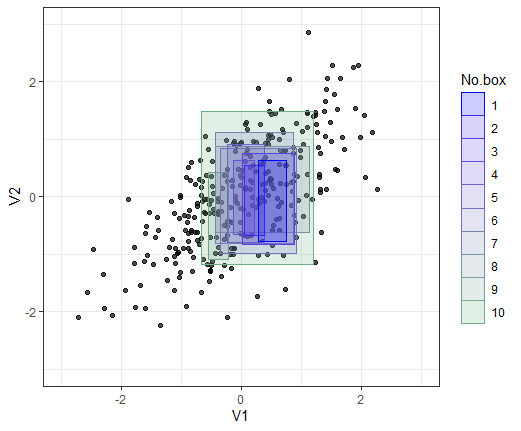}
\caption{Region obtained by PRIM.}
\label{Fig.prim}
\end{figure}

\begin{figure}[!t]
\centering
%\begin{varwidth}{\linewidth}
\subfloat[Principal - PRIM]{\includegraphics[width=4cm,height=4cm]{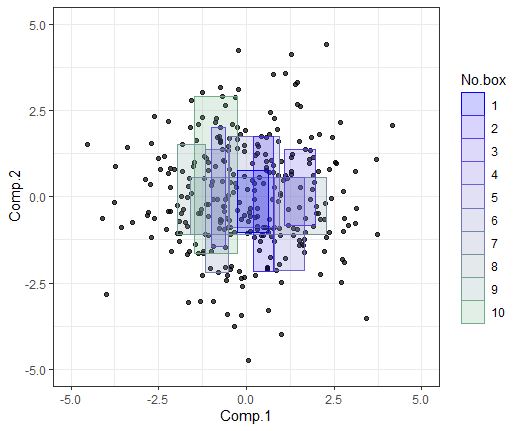}}
\subfloat[Pettiest - PRIM]{\includegraphics[width=4cm,height=4cm]{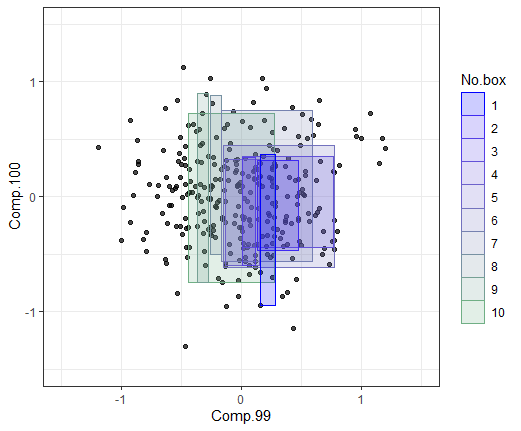}}\\
\subfloat[Principal - fastPRIM]{\includegraphics[width=4cm,height=4cm]{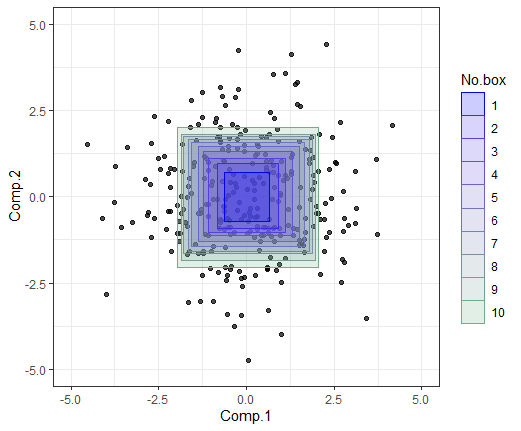}}
\subfloat[Pettiest - fastPRIM]{\includegraphics[width=4cm,height=4cm]{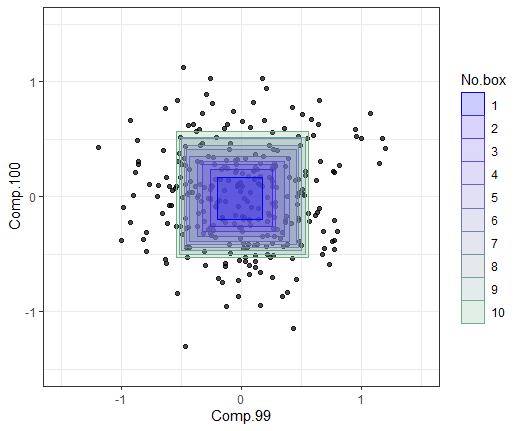}}\\
%\end{varwidth}
\caption{Simulation results by method.} %The left column shows principal components; the right column shows pettiest components. The top row shows PRIM; the bottom row shows fastPRIM.
\label{Fig.main}
\end{figure}

We generated a dataset with 300 observations and 100 dimensions, which follows the multivariate normal distribution $\mathcal{N}_p (0, \Sigma)$. The covariance matrix $\Sigma$ is assumed to have its first two diagonal elements as 1, and cov($x_1,x_2$)$= 0.7$; the last two diagonal elements are 12 with cov($x_{99}, x_{100}$)$= 8$; the remaining diagonal elements are 6 and all other places are 0. A pure PRIM, no rotation nor reduction, with ten iterations of covering, was then applied for contrast purposes. The result is shown in Fig. \ref{Fig.prim} for the first two dimensions. We next standardized the dataset to use the correlation matrix and apply the rotation in the direction of the eigenvectors. Once the rotation was performed, the two new variables with larger variances are the principal components, and the two variables with smaller variances are the pettiest components. The next step was to run both PRIM and fastPRIM on the two principal components and the two pettiest components, both with $\beta = 0.1$. The results are shown in Fig. \ref{Fig.main}. Each box results from a whole peeling period and the order of the boxes after ten stages of covering is shown by grading color. The fastPRIM algorithm exhibited nested rectangular boxes in contrast with the PRIM's messy ones. There is a striking difference in the result of both algorithms (PRIM and fastPRIM) between principal and pettiest components: the final boxes obtained by fastPRIM are almost five times smaller than those of its competitor.

Table \ref{table.main} shows the quantitative results in terms of the empirical density over volume of the box. Looking at it by columns, we can compare the different methods under the same covering step. First, the classical PRIM results in too small densities, thus rendering it completely useless. This in the end comes as no surprise considering the curse of dimensionality. In this sense, Table \ref{table.main} also reveals the benefits of dimension reduction, since any of the other options will do much better. That is, when using principal or pettiest components analysis to reduce dimensionality, from 100 dimensions to 2 in our example, we get much better densities. In fact, fastPRIM improves the densities even more. However, the optimal improvement comes when the pettiest components are used; in this case, the densities show an increase of an order of magnitude. Finally, looking at the table by rows, one can track the variations of the densities during covering procedures. There is an interesting contrast in that for PRIM the density starts from a really small value and then increases monotonically; whereas for all other cases, a final decrease is accomplished as expected, though it is not monotonic. 

Table \ref{table.stat} shows empirical bias and variance for the different methods. The bias was calculated as the distance between the mean value and the origin. The results were obtained by averaging over 100 simulations from the same distribution used before. We can see that the variance of PRIM is too high, compared to the other mechanisms with the dimension reduction, corresponding to what we would expect. Also, PRIM with principal components and fastPRIM with principal components have variances that are close, but there is a huge reduction of bias between the two methods. The same is true for PRIM and fastPRIM both with pettiest components. This illustrates the superiority of fastPRIM in this scenario. Finally, notice that pettiest components is producing a tenfold decrease in variance, whether we are looking at PRIM or fastPRIM; the bias also decreases. The final outcome is that fastPRIM with pettiest components, having lowest variance and bias among all the strategies, minimizes the empirical mean squared error.

\begin{table*}[!t]
\caption{Density of boxes per volume by different method}{
\resizebox{\textwidth}{!}{
  \begin{tabular}{lcccccccccc}
    %\toprule
     & \textbf{1} & \textbf{2} & \textbf{3} & \textbf{4} & \textbf{5} & \textbf{6} & \textbf{7} & \textbf{8} & \textbf{9} & \textbf{10}\\
    %\hline
    PRIM & 1.07e-73 & 1.57e-73 & 1.76e-73 & 2.14e-73 & 2.59e-73 & 3.04e-73 & 3.45e-73 & 3.87e-73 & 4.26e-73 & 4.65e-73\\
    %\hline
    PRIM-Principal & 15.7 & 16.7 & 14.2 & 14.7 & 14.3 & 14.2 & 14.0 & 13.9 & 13.8 & 13.0\\
    %\hline
    fastPRIM-Principal & 16.3 & 15.6 & 16.2 & 16.3 & 14.5 & 13.2 & 13.1 & 13.1 & 12.7 & 11.8\\
    %\hline
    PRIM-Pettiest & 182 & 168 & 187 & 152 & 155 & 142 & 130 & 132 & 132 & 127\\
    %\hline
    fastPRIM-Pettiest & 215 & 240 & 237 & 218 & 207 & 186 & 189 & 177 & 169 & 161
    %\bottomrule
  \end{tabular}}}
\label{table.main}
\end{table*}

\begin{table}[!t]
\tiny
\centering
\caption{Empirical variance and bias by method}{
%\resizebox{\textwidth}{!}{
  \begin{tabular}{lcccccccccc}
     & PRIM & PRIM-Principal & PRIM-Pettiest & fastPRIM-Principal & fastPRIM-Pettiest\\
    \textbf{Variance} & 104 & 2.28  & 0.204 & 2.23 & 0.149\\
    \textbf{Bias} & 0.310 & 0.145 & 0.1 & 0.00579  & 0.000653
  \end{tabular}}%}
\label{table.stat}
\end{table}

\section{Example: MNIST Dataset}\label{Sec:MNIST}

MNIST is a famous dataset of handwritten digits widely used in machine learning \cite{LeCunEtAl1999}. Using principal components analysis with the MNIST dataset is somewhat common. In fact, a few principal components have been used to reconstruct the digits \cite[pp.~433-445]{VanderPlas2016}. With this idea in mind, we suggest that the platonic pattern of each digit (the archetypical representation of a digit) is close to the mode.  Therefore, here we use pettiest components and show that for PRIM and fastPRIM applied to MNIST, the pettiest components give a higher active information than the principal components. Thus, in words, the machine can use principal components in order to \textit{read} a digit, learning all its variability (see \cite[pp.~536-539]{HastieTibshiraniFriedman2009}, though with a different hand-written digits dataset). But if the machine is going to learn to \textit{write} it, it is better to go for the mode in pettiest components. Thus it is better to work with pettiest components in order to generate the actual image.

\begin{figure}[!t]
\centering 
\includegraphics[width=0.5\textwidth,height=4cm]{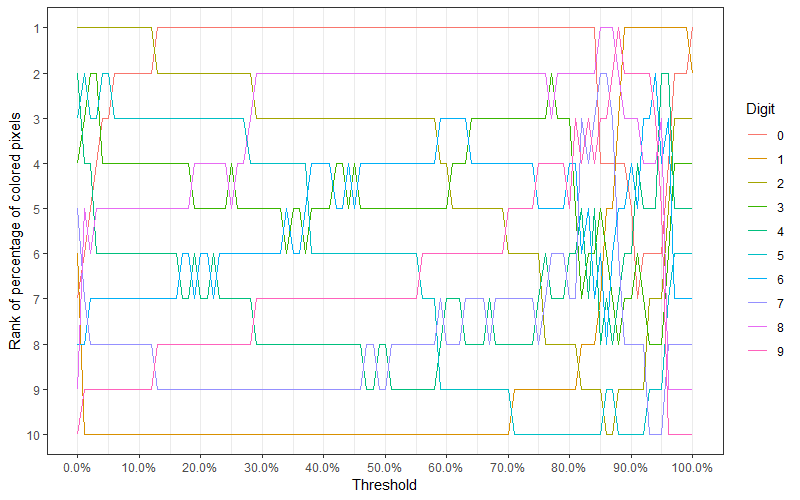}
\caption{Ranking changes by threshold}\label{Fig.rank}
\end{figure}

Since our goal is the identification of modal patterns for digits, only the training dataset is used. The data consist of grey levels, from white to black, of $28 \times 28$ pixels for 60,000 observations. This results in a $60,000 \times 784$ matrix. All images are centered by the center of mass of the pixels. We first split the big dataset by digit to get 10 smaller datasets with size near $6000 \times 784$. Notice that the graphs are comprised mostly of white pixels, corresponding to zero value. If these white pixels are not removed, the modes will obviously be biased towards those values. Therefore these zero points make mode hunting strategies unsuitable. We need to find a threshold to make sure that most observations will be colored on those pixels. The threshold will cause a reduction in dimensionality, which should have different degrees corresponding to different numbers. The relative ranking of these reductions should be stable when the threshold changes. So we measure this reduction by the percentage of pixels we choose to keep and rank it by number from high to low. The ranking plot by threshold is shown in Fig. \ref{Fig.rank}, from which it is observed that the ranking is relatively stable when the threshold is lower than 60\%. 

\begin{figure}[!t]
\centering

\subfloat[Cross-validation for 0]{\includegraphics[width=4cm,height=2.5cm]{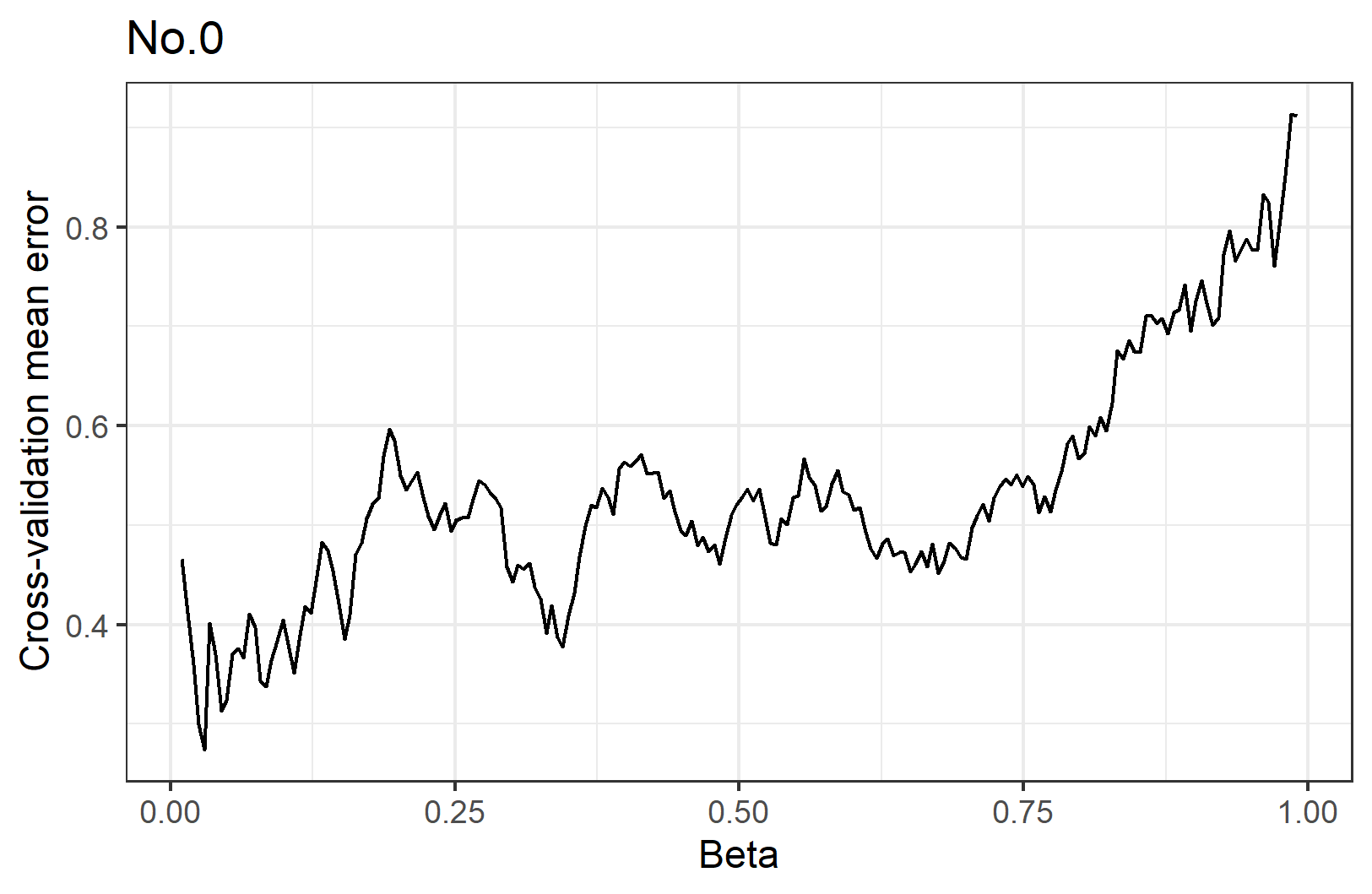}}
\subfloat[Cross-validation for 1]{\includegraphics[width=4cm,height=2.5cm]{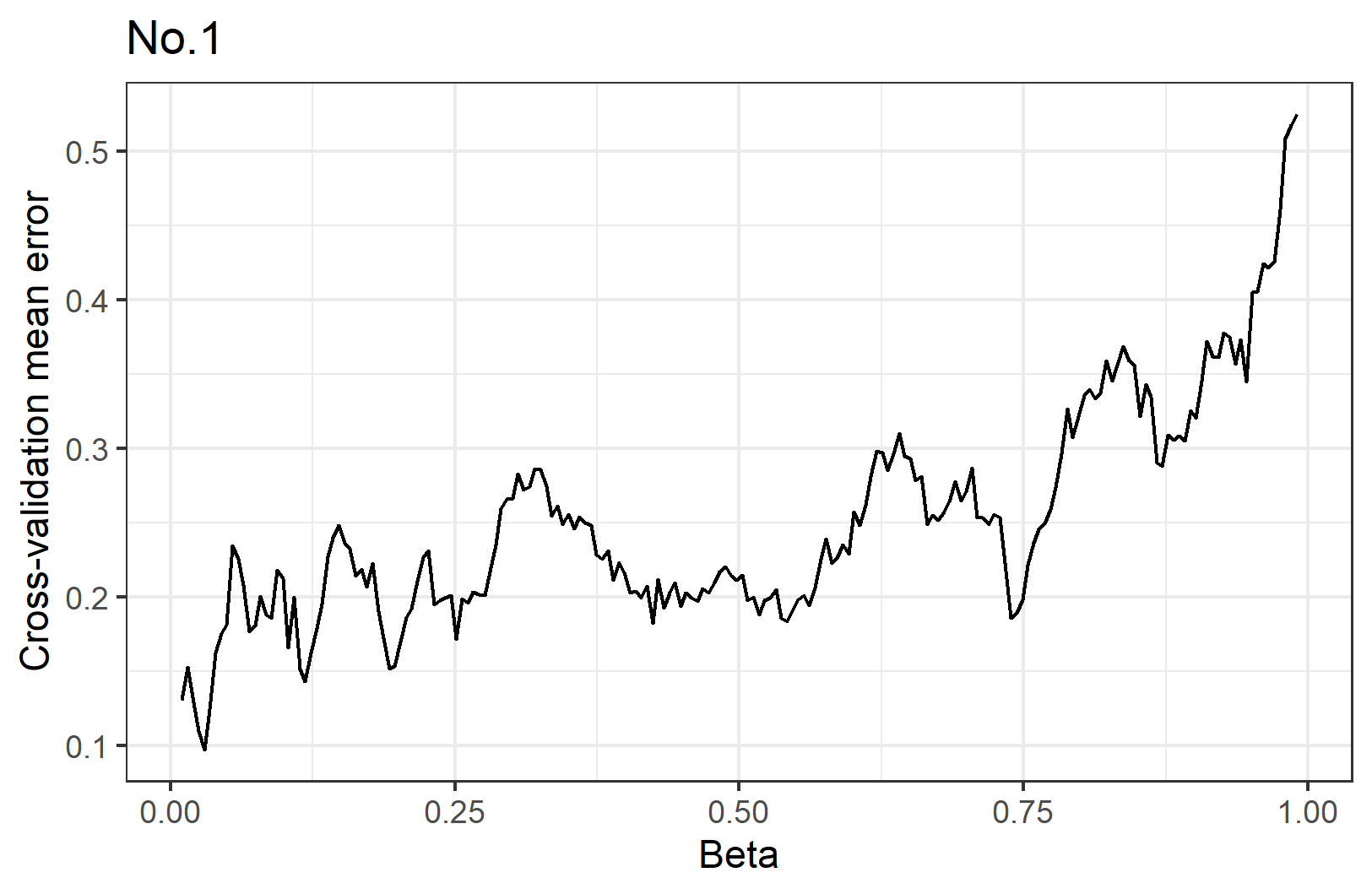}}

\subfloat[Cross-validation for 2]{\includegraphics[width=4cm,height=2.5cm]{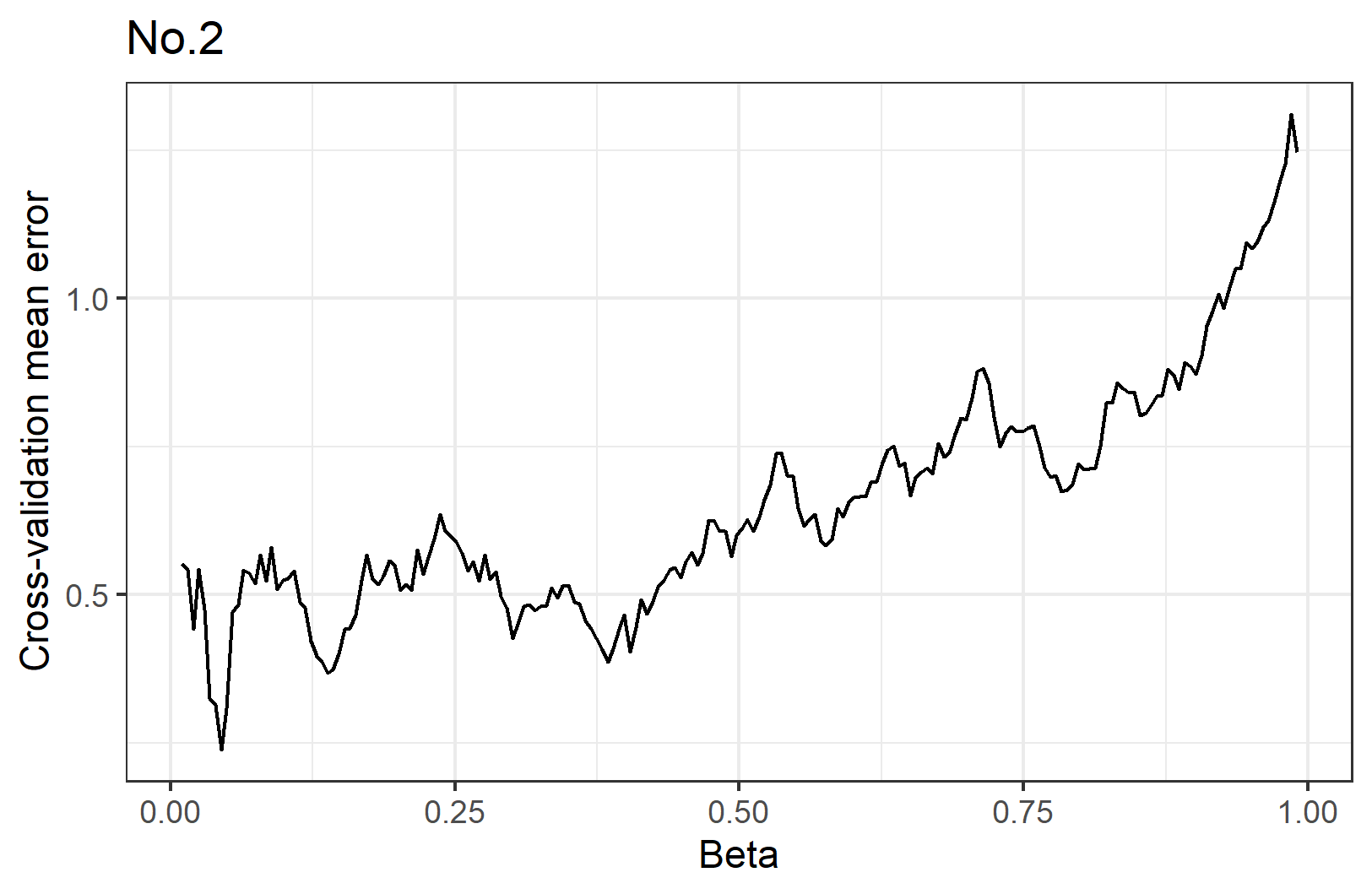}}
\subfloat[Cross-validation for 3]{\includegraphics[width=4cm,height=2.5cm]{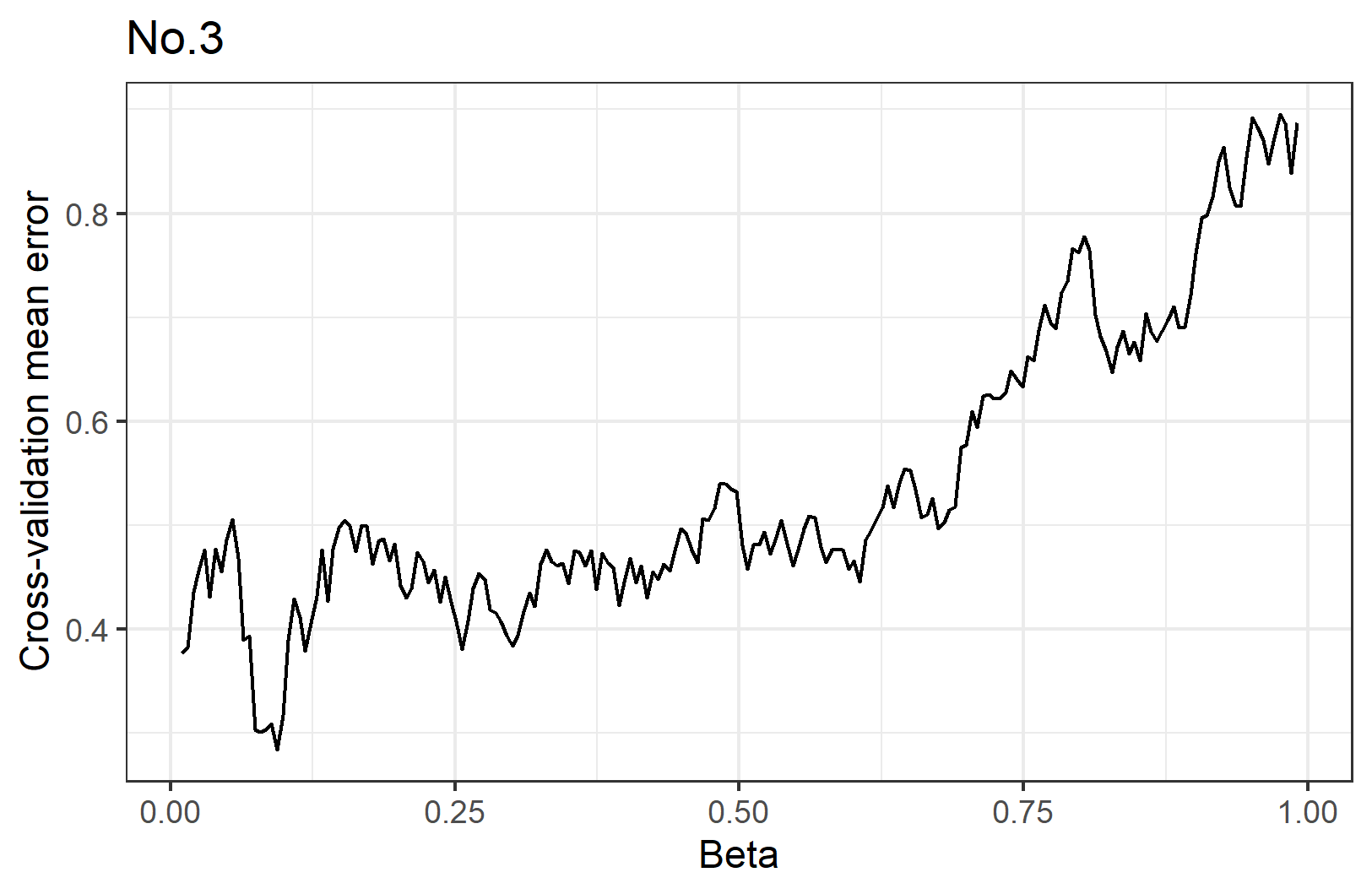}}

\subfloat[Cross-validation for 4]{\includegraphics[width=4cm,height=2.5cm]{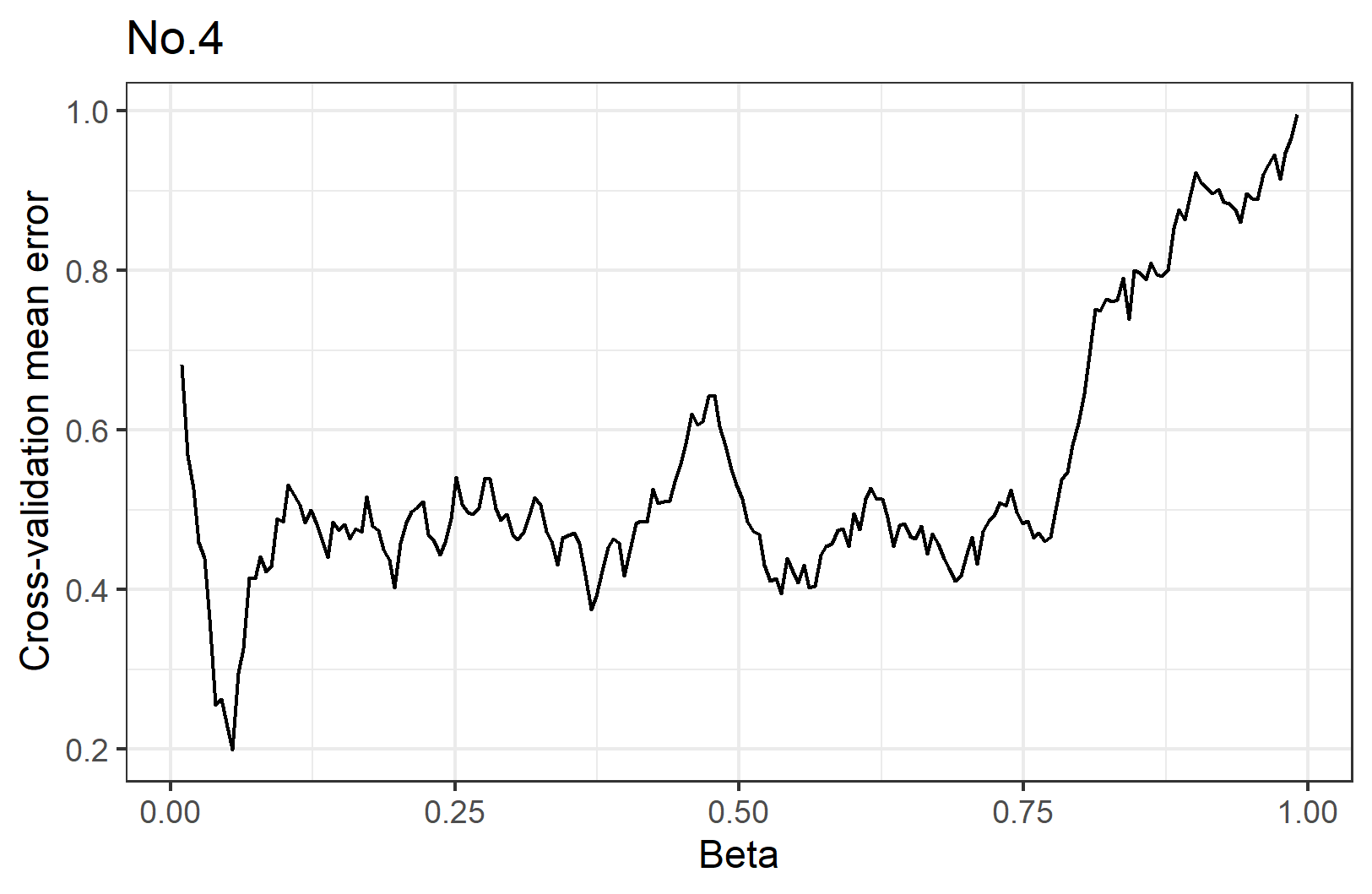}}
\subfloat[Cross-validation for 5]{\includegraphics[width=4cm,height=2.5cm]{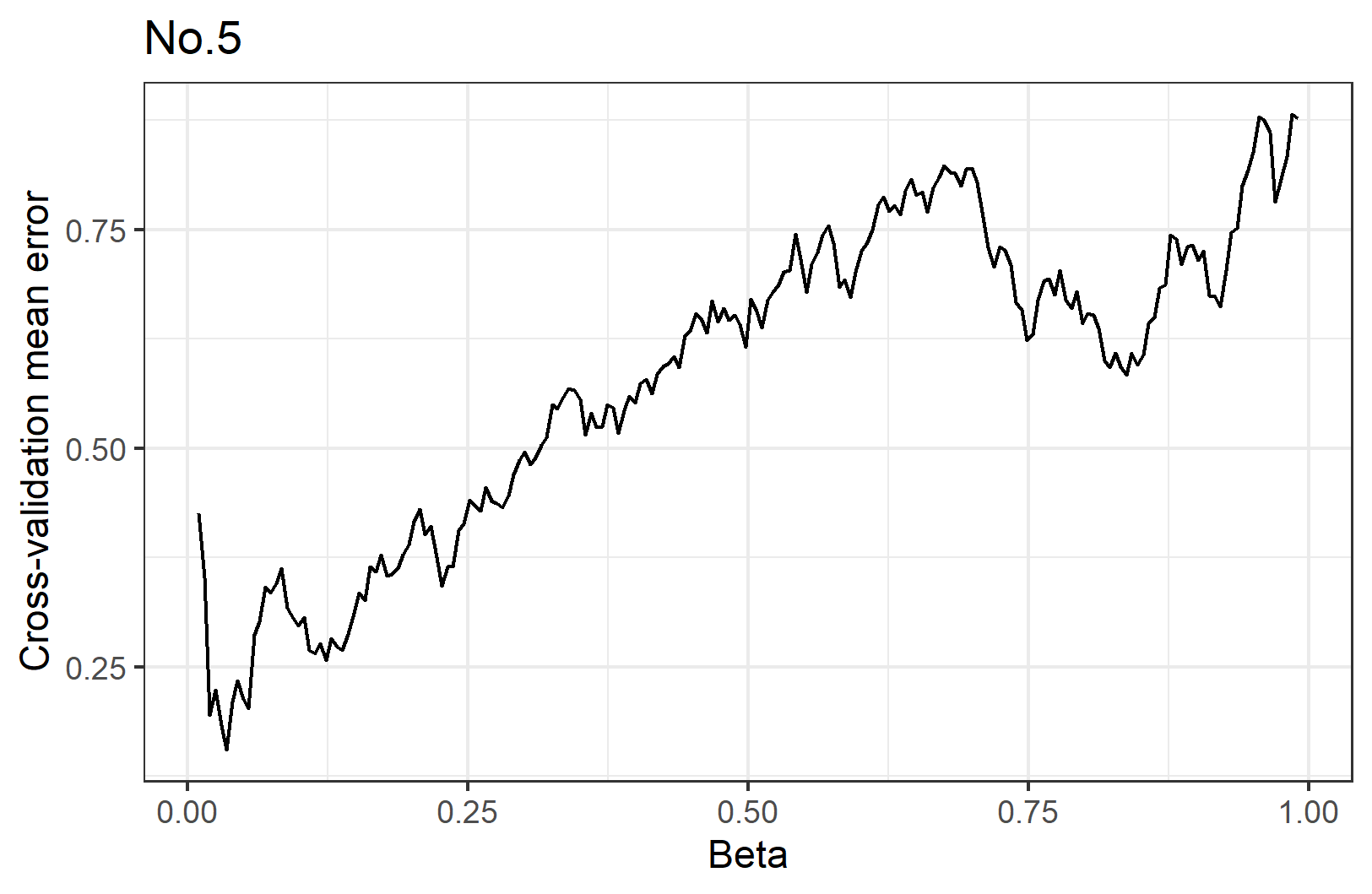}}

\subfloat[Cross-validation for 6]{\includegraphics[width=4cm,height=2.5cm]{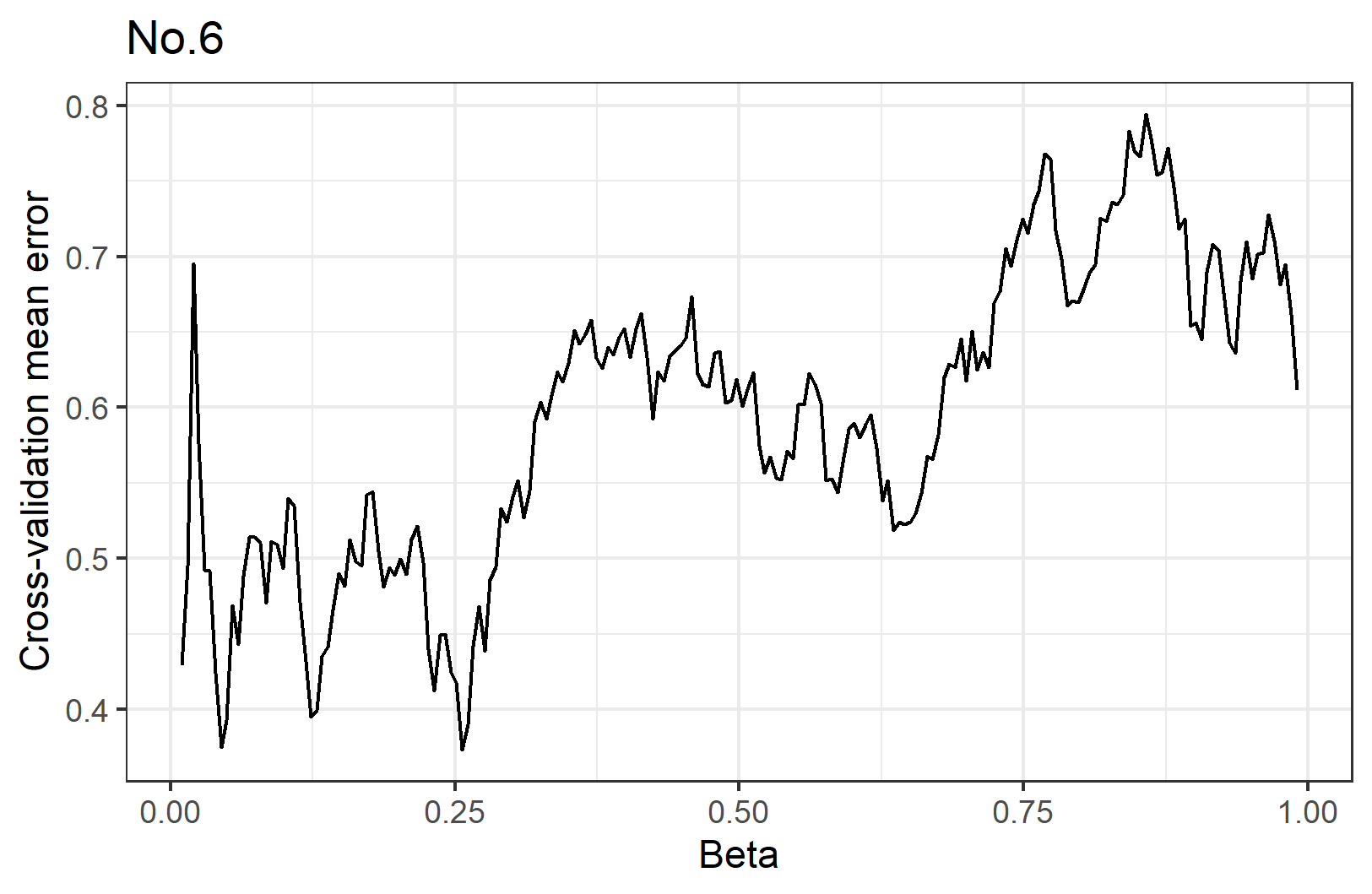}}
\subfloat[Cross-validation for 7]{\includegraphics[width=4cm,height=2.5cm]{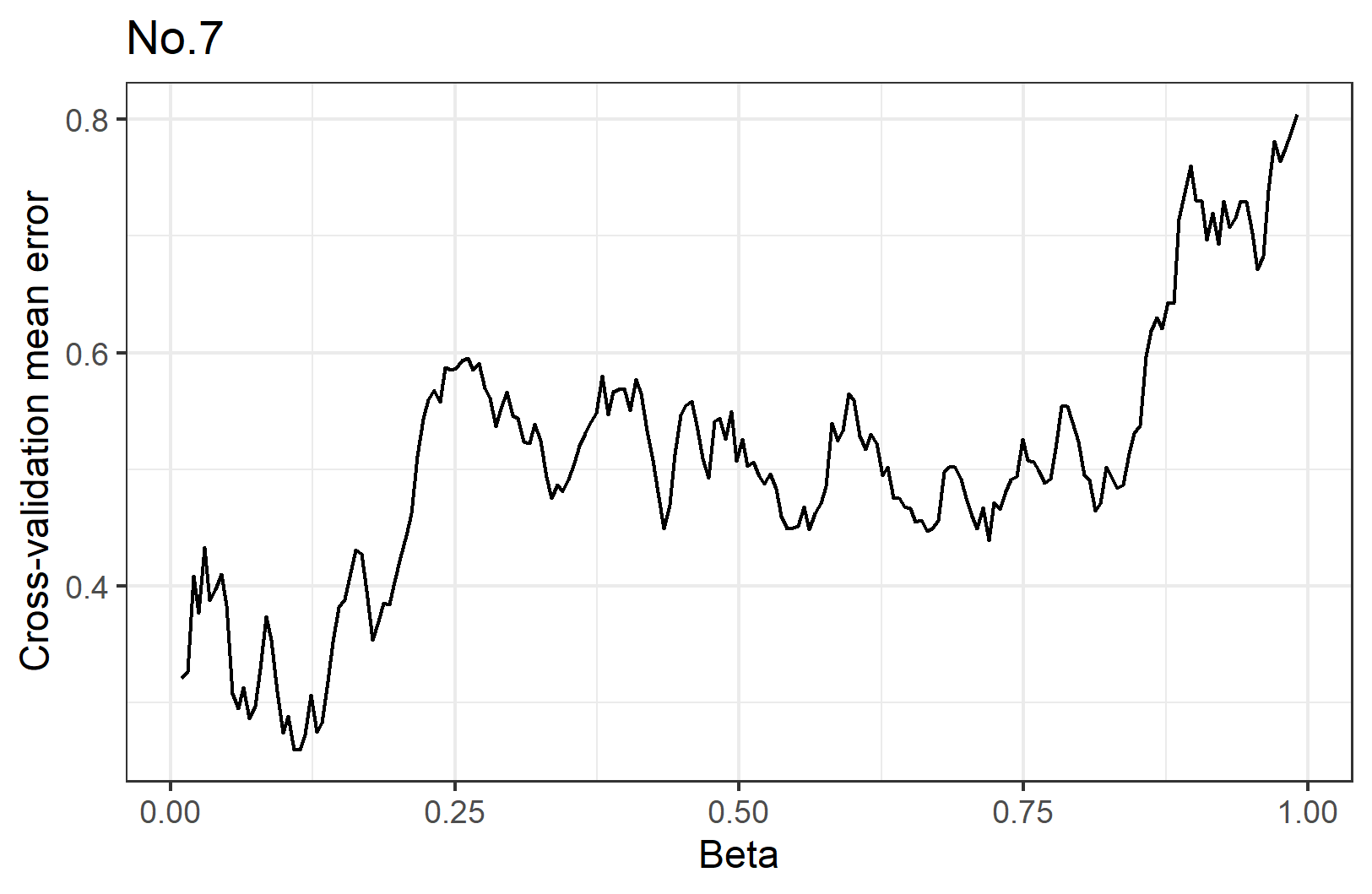}}

\subfloat[Cross-validation for 8]{\includegraphics[width=4cm,height=2.5cm]{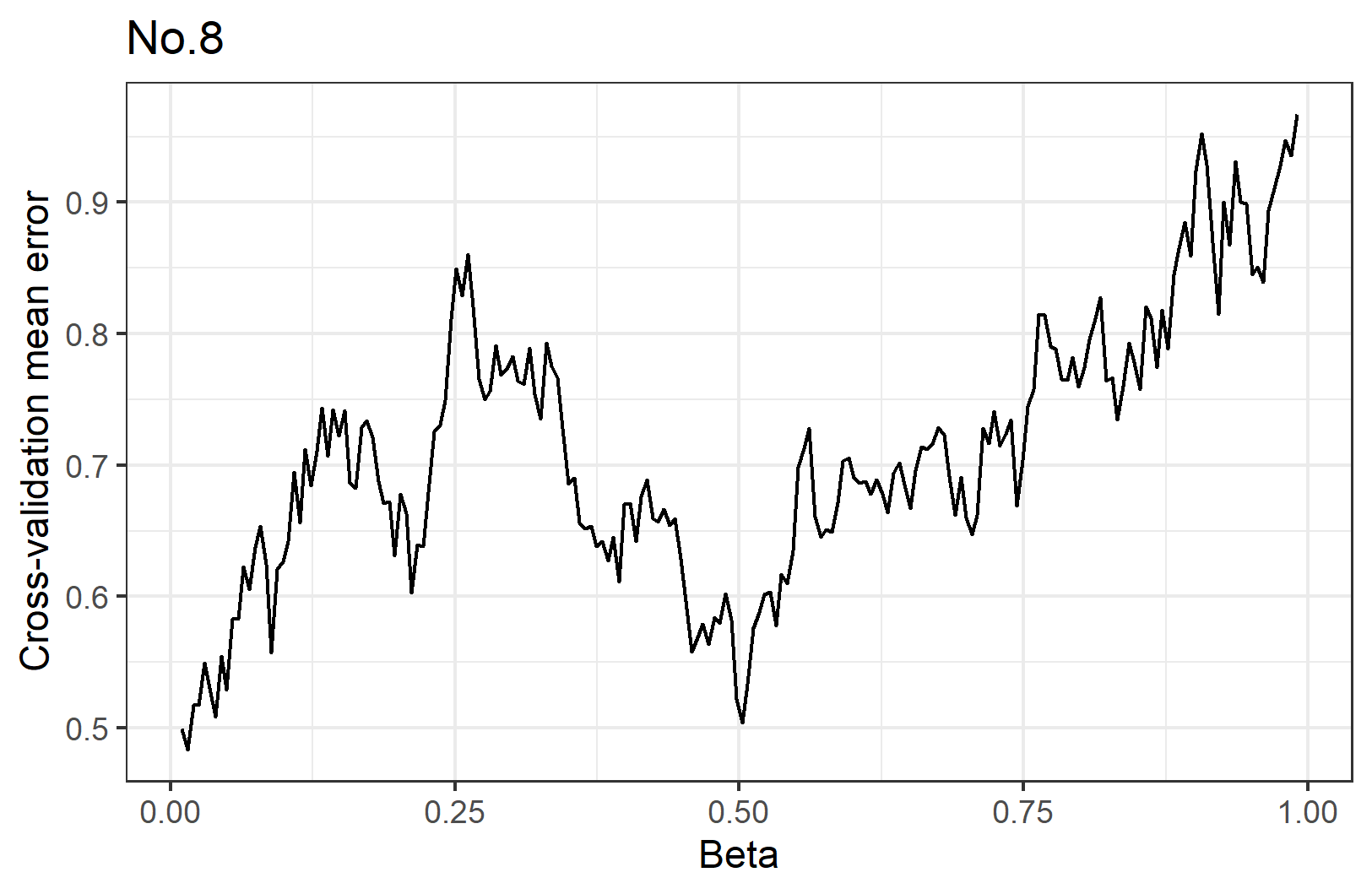}}
\subfloat[Cross-validation for 9]{\includegraphics[width=4cm,height=2.5cm]{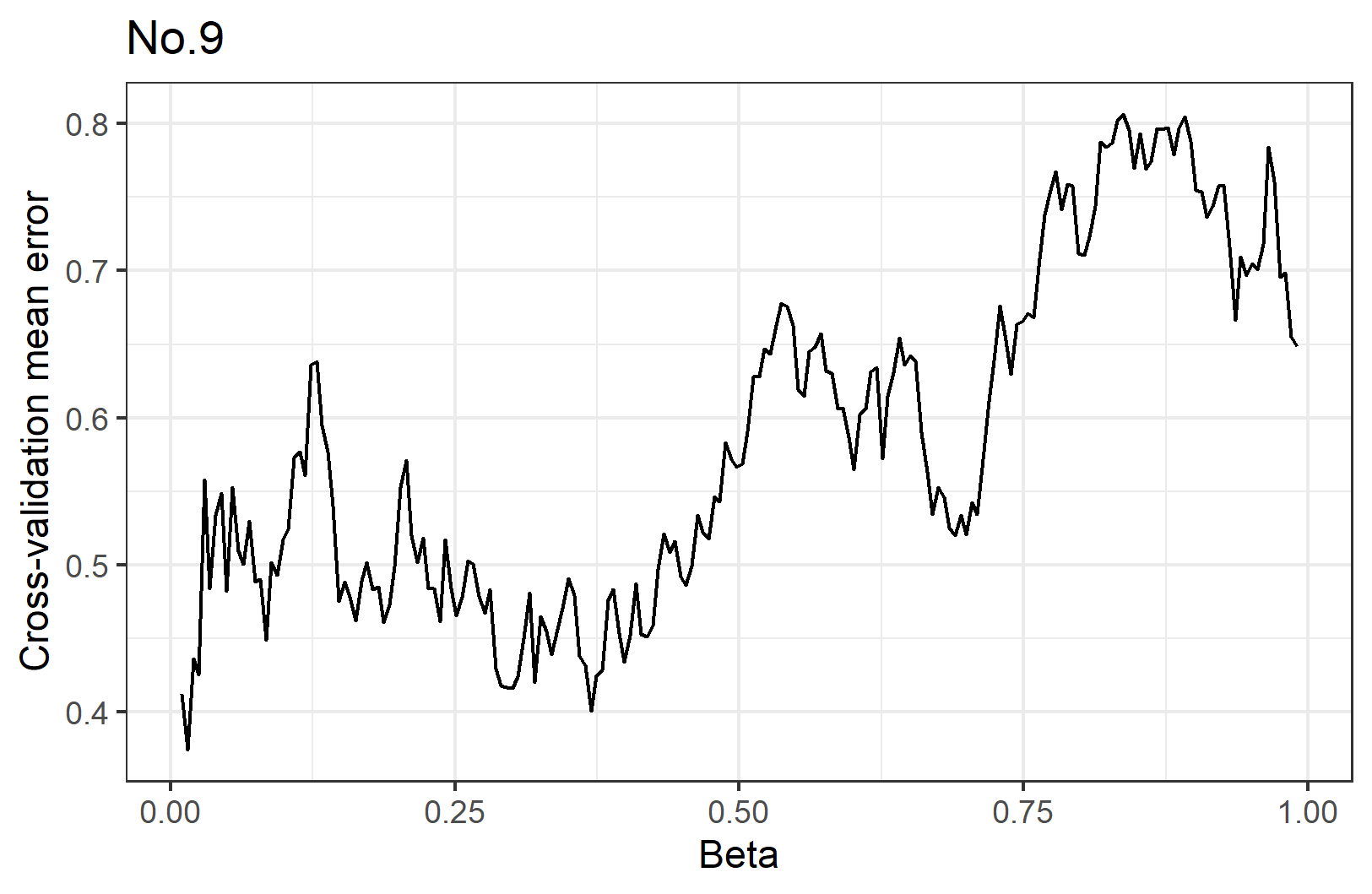}}

\caption{Cross-validation per digit}
\label{Fig.cv}
\end{figure}

In this fashion, we obtain 10 datasets corresponding to each digit, with about 6000 observations and dimensionality smaller than 784. In order to apply PRIM and fastPRIM with principal and pettiest components analysis on each of the 10 datasets, we need first to determine for each digit the right probability $\beta$ of the region that will contain the mode. We achieve this by minimizing the mean squared error through a 10-fold cross-validation, digit by digit for fastPRIM with pettiest components (Fig. \ref{Fig.cv}). Then, in oder to be able to make comparisons between strategies, we use the same $\beta$-optimized value for fastPRIM with principal components, as well as for PRIM with pettiest and principal components.  When the size of $\beta$ permits it, we have several iterations of the covering process for PRIM, making $\beta'$ the probability of the box in each iteration of the covering process, to obtain a final region of probability $\beta= 1 - (1 - \beta')^t$ after $t$ iterations  (Table \ref{crossv}). Thus, for instance, the digits 0 and 1 have just a single stage of covering, whereas the digits 8 and 9 have 7 and 4, respectively. The results are shown in Figs.\,\ref{Fig.example1} and \ref{Fig.example2}, where the superiority of pettiest over principal components in detecting the modes is clearly observable.

\begin{table*}[!t]
\scriptsize
\caption{Final region size and iterations per digit}\label{crossv}%{
%\resizebox{\textwidth}{!}{
\centering
  \begin{tabular}{ccccccccccc}
    \  & 0 & 1 & 2 & 3 & 4  & 5 & 6 & 7 & 8 & 9 \\
    $\beta$ & 2.96\% & 3.21\% & 4.45\% & 9.37\% & 5.43\% & 3.46\% & 25.6\% & 11.3\% & 50.2\% & 36.9\%\\
	iterations & 1 & 1 & 1 & 1 & 1 & 1 & 3 & 1 & 7 & 4
  \end{tabular}%}}
\end{table*}

\begin{figure}[!t]
\centering

\subfloat[Principal with fastPRIM for 0]{\includegraphics[width=4cm,height=3cm]{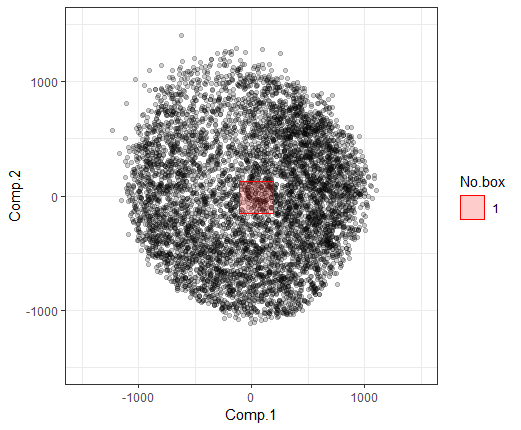}}
\subfloat[Pettiest with fastPRIM for 0]{\includegraphics[width=4cm,height=3cm]{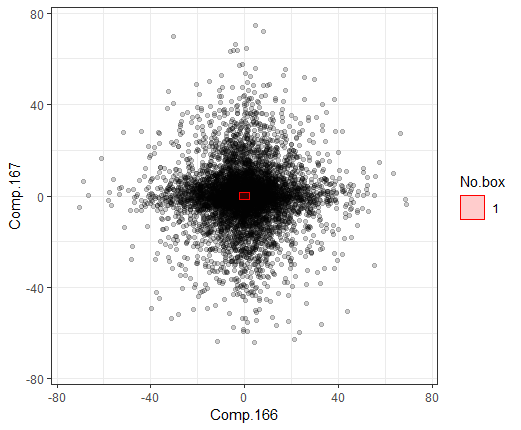}}

\subfloat[Principal with fastPRIM for 1]{\includegraphics[width=4cm,height=3cm]{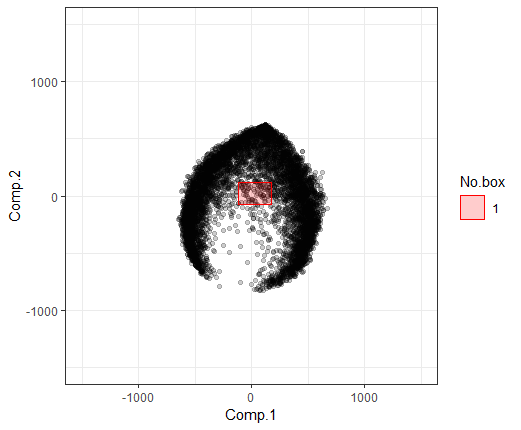}}
\subfloat[Pettiest  with fastPRIM for 1]{\includegraphics[width=4cm,height=3cm]{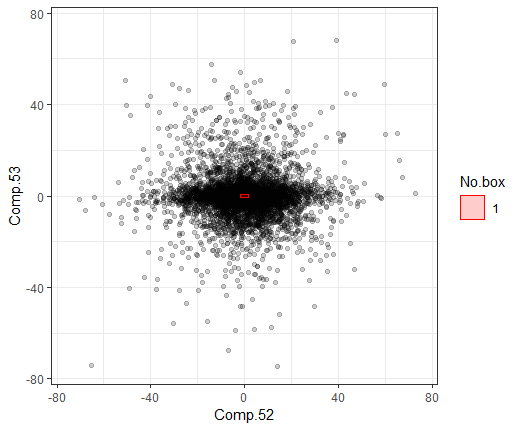}}

\subfloat[Principal with fastPRIM for 2]{\includegraphics[width=4cm,height=3cm]{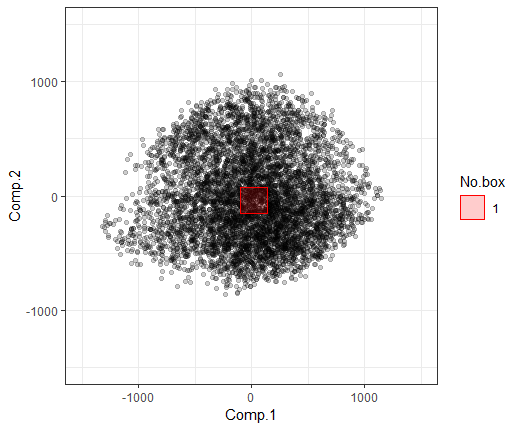}}
\subfloat[Pettiest with fastPRIM for 2]{\includegraphics[width=4cm,height=3cm]{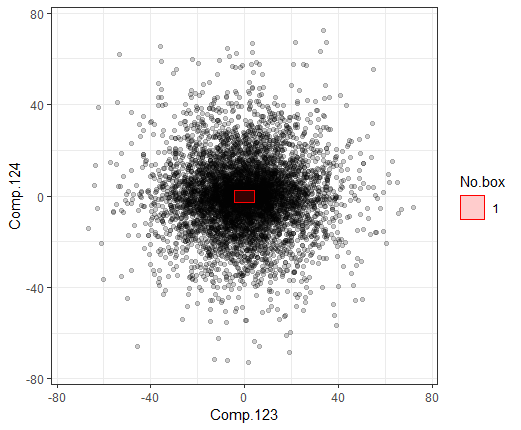}}

\subfloat[Principal with fastPRIM for 3]{\includegraphics[width=4cm,height=3cm]{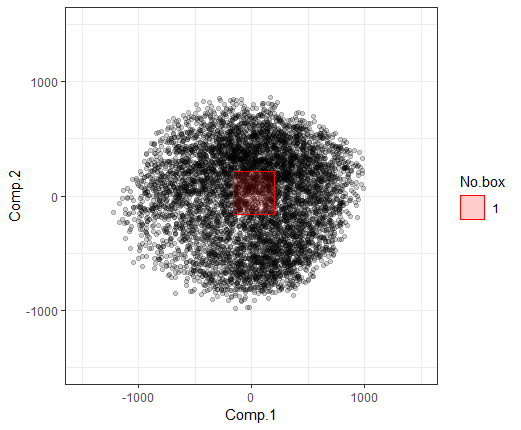}}
\subfloat[Pettiest with fastPRIM for 3]{\includegraphics[width=4cm,height=3cm]{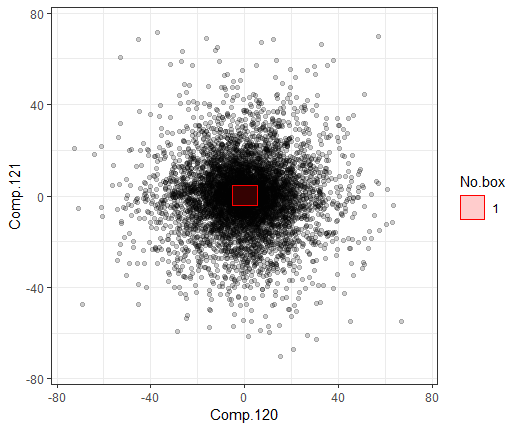}}

\subfloat[Principal with fastPRIM for 4]{\includegraphics[width=4cm,height=3cm]{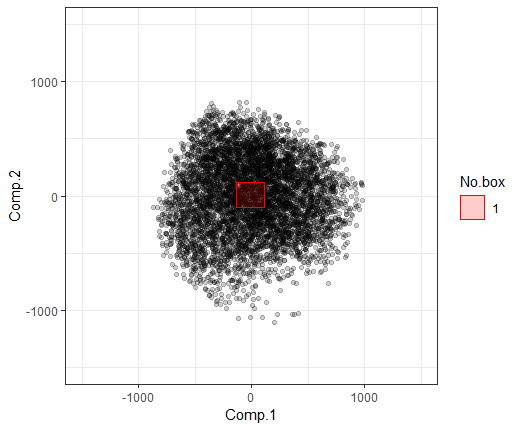}}
\subfloat[Pettiest with fastPRIM for 4]{\includegraphics[width=4cm,height=3cm]{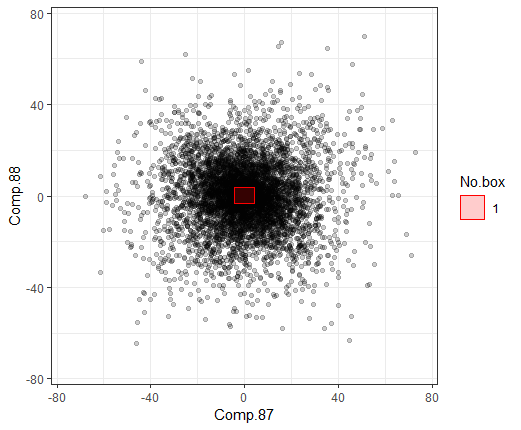}}

\caption{MNIST modeling results 0--4 (fastPRIM)}
\label{Fig.example1}
\end{figure}

%\clearpage
\begin{figure}[!t]\ContinuedFloat
\centering
\subfloat[Principal with fastPRIM for 5]{\includegraphics[width=4cm,height=3cm]{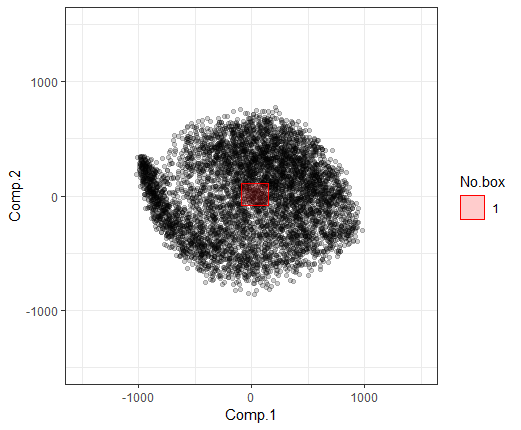}}
\subfloat[Pettiest with fastPRIM for 5]{\includegraphics[width=4cm,height=3cm]{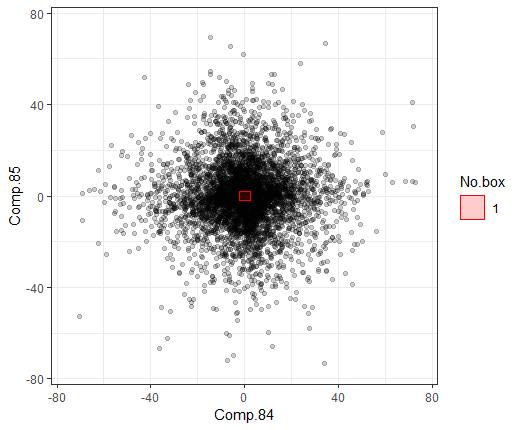}}

\subfloat[Principal with fastPRIM for 6]{\includegraphics[width=4cm,height=3cm]{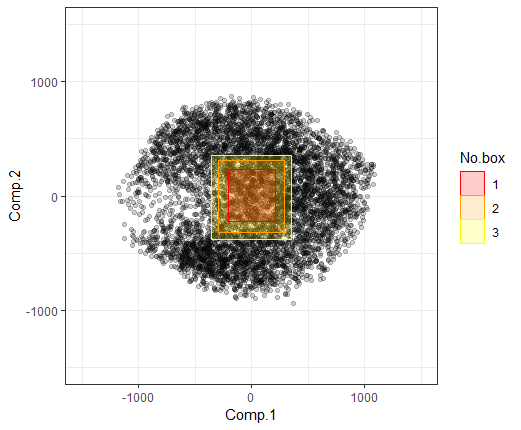}}
\subfloat[Pettiest with fastPRIM for 6]{\includegraphics[width=4cm,height=3cm]{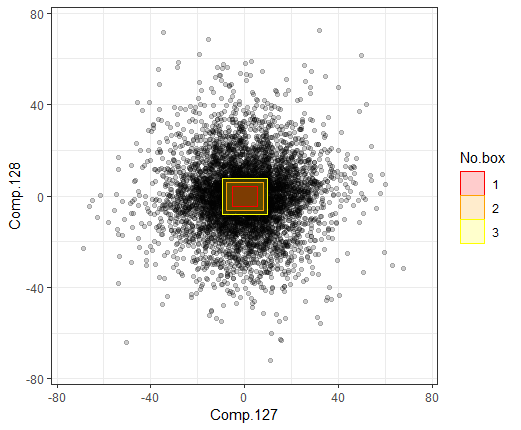}}

\subfloat[Principal with fastPRIM for 7]{\includegraphics[width=4cm,height=3cm]{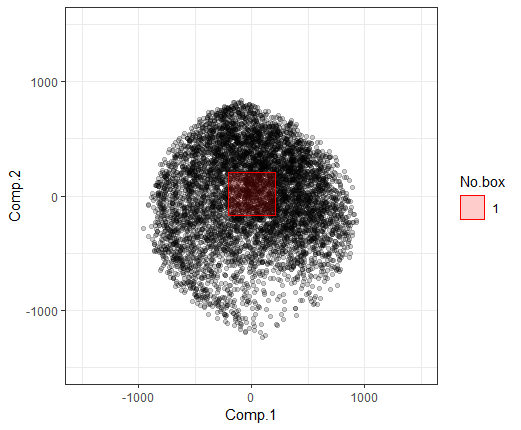}}
\subfloat[Pettiest with fastPRIM for 7]{\includegraphics[width=4cm,height=3cm]{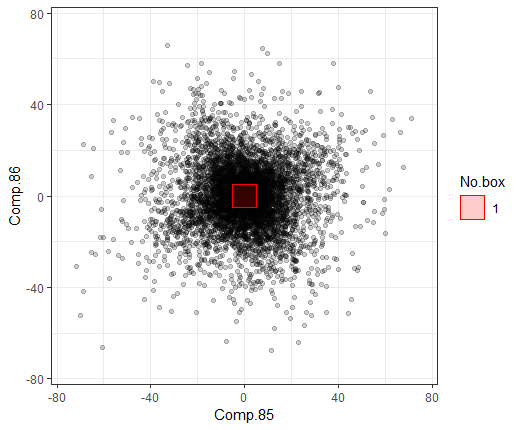}}

\subfloat[Principal with fastPRIM for 8]{\includegraphics[width=4cm,height=3cm]{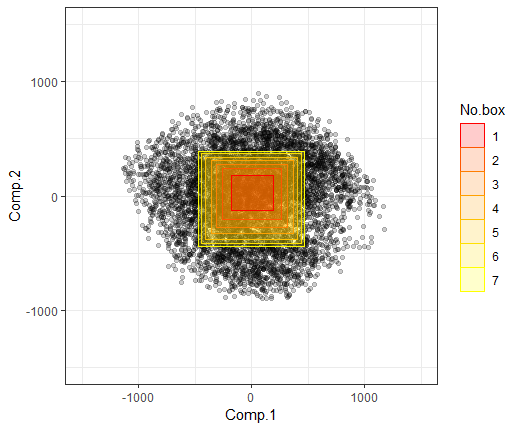}}
\subfloat[Pettiest with fastPRIM for 8]{\includegraphics[width=4cm,height=3cm]{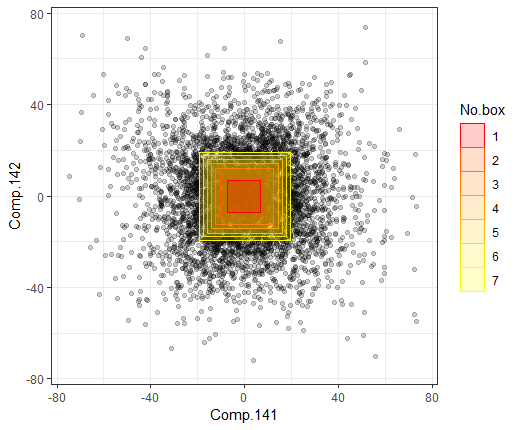}}

\subfloat[Principal with fastPRIM for 9]{\includegraphics[width=4cm,height=3cm]{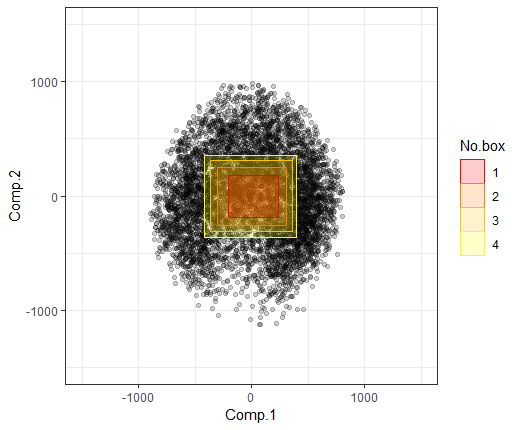}}
\subfloat[Pettiest with fastPRIM for 9]{\includegraphics[width=4cm,height=3cm]{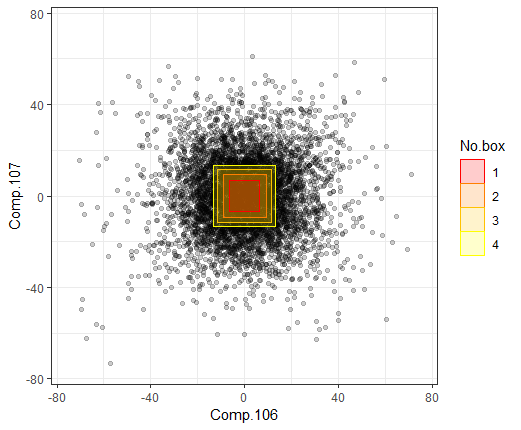}}
\caption{MNIST modeling results 5--9 (fastPRIM)}
\end{figure}

\begin{figure}[!t]
\centering

\subfloat[Principal  with PRIM for 0]{\includegraphics[width=4cm,height=3cm]{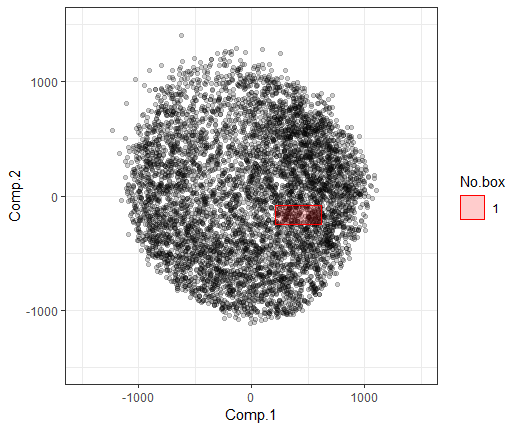}}
\subfloat[Pettiest  with PRIM for 0]{\includegraphics[width=4cm,height=3cm]{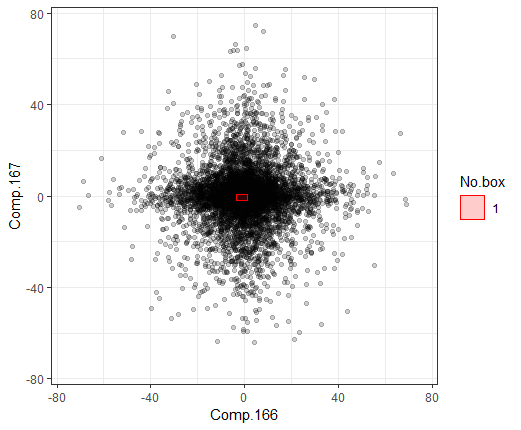}}

\subfloat[Principal  with PRIM for 1]{\includegraphics[width=4cm,height=3cm]{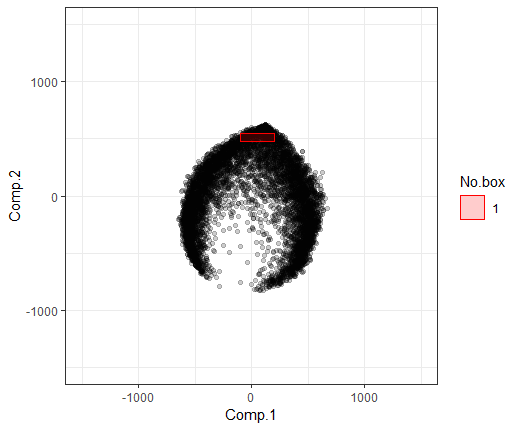}}
\subfloat[Pettiest  with PRIM for 1]{\includegraphics[width=4cm,height=3cm]{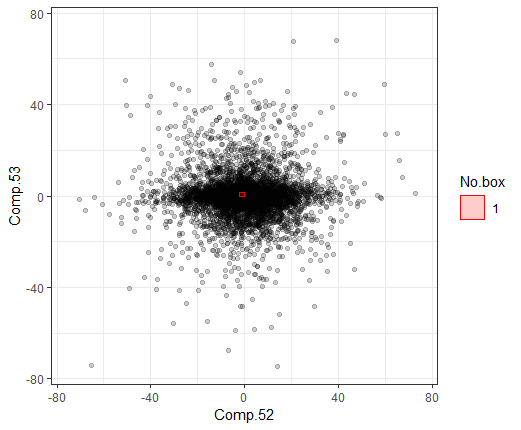}}

\subfloat[Principal  with PRIM for 2]{\includegraphics[width=4cm,height=3cm]{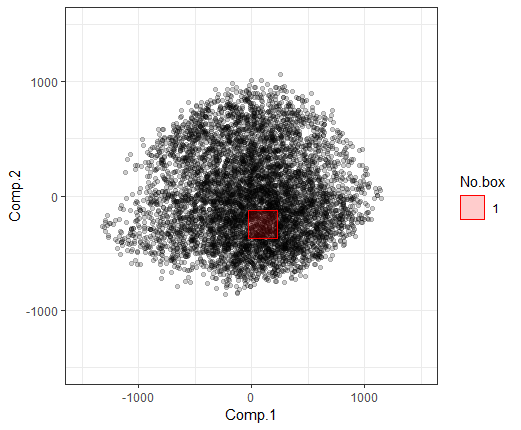}}
\subfloat[Pettiest  with PRIM for 2]{\includegraphics[width=4cm,height=3cm]{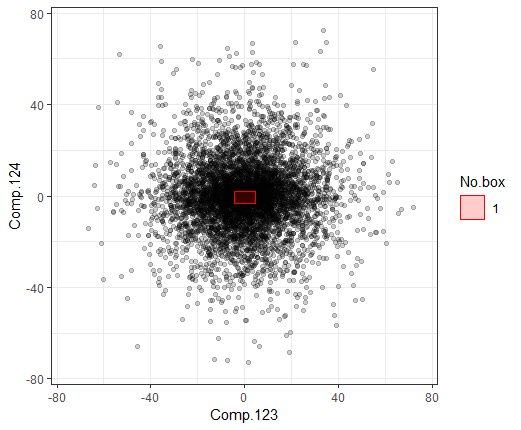}}

\subfloat[Principal  with PRIM for 3]{\includegraphics[width=4cm,height=3cm]{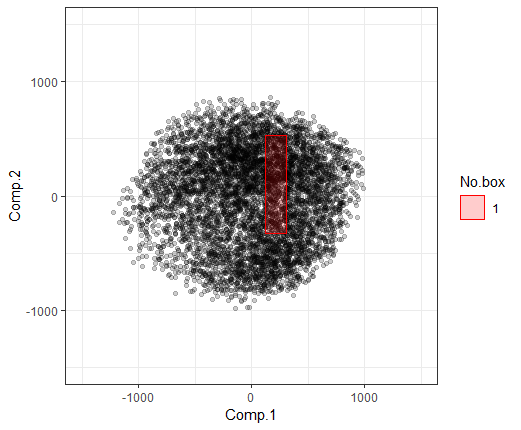}}
\subfloat[Pettiest  with PRIM for 3]{\includegraphics[width=4cm,height=3cm]{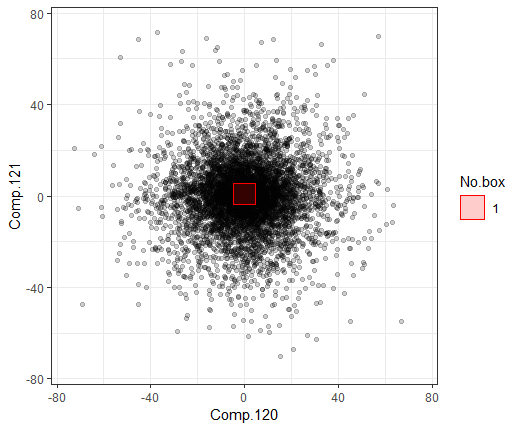}}

\subfloat[Principal  with PRIM for 4]{\includegraphics[width=4cm,height=3cm]{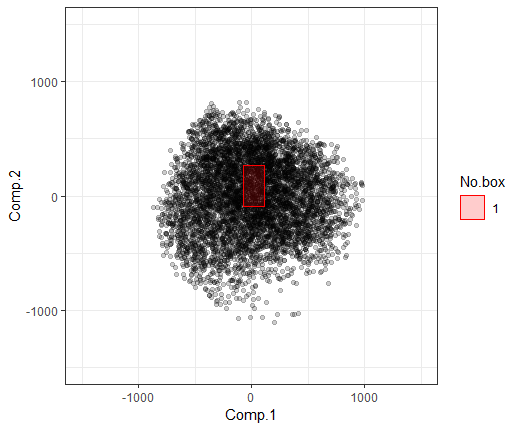}}
\subfloat[Pettiest  with PRIM for 4]{\includegraphics[width=4cm,height=3cm]{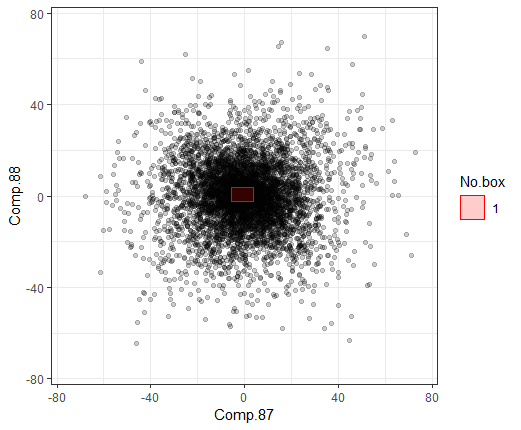}}

\caption{MNIST modeling results 0--4 (PRIM)}
\label{Fig.example2}
\end{figure}

%\clearpage
\begin{figure}[!t]\ContinuedFloat
\centering

\subfloat[Principal  with PRIM for 5]{\includegraphics[width=4cm,height=3cm]{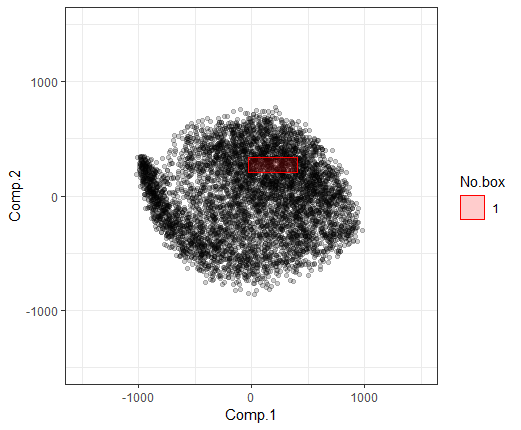}}
\subfloat[Pettiest  with PRIM for 5]{\includegraphics[width=4cm,height=3cm]{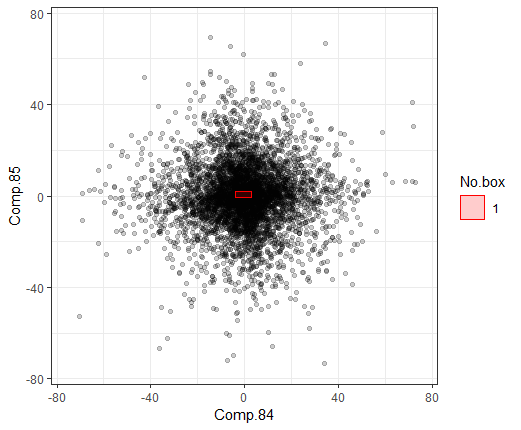}}

\subfloat[Principal  with PRIM for 6]{\includegraphics[width=4cm,height=3cm]{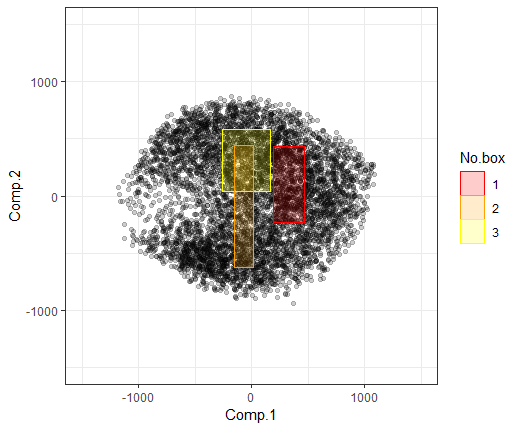}}
\subfloat[Pettiest  with PRIM for 6]{\includegraphics[width=4cm,height=3cm]{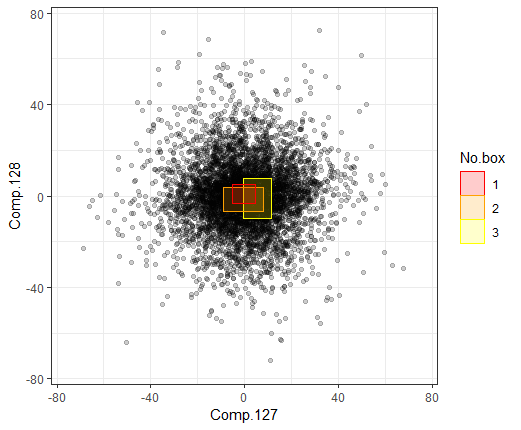}}

\subfloat[Principal  with PRIM for 7]{\includegraphics[width=4cm,height=3cm]{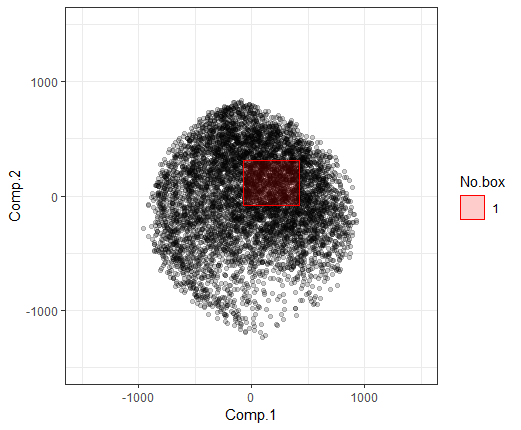}}
\subfloat[Pettiest  with PRIM for 7]{\includegraphics[width=4cm,height=3cm]{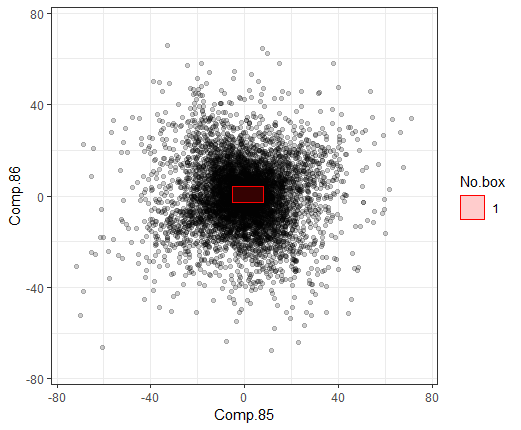}}

\subfloat[Principal  with PRIM for 8]{\includegraphics[width=4cm,height=3cm]{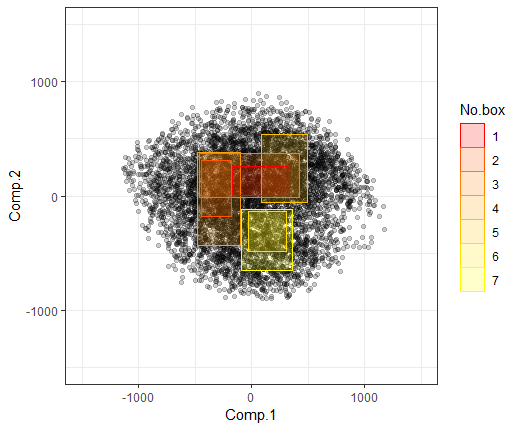}}
\subfloat[Pettiest  with PRIM for 8]{\includegraphics[width=4cm,height=3cm]{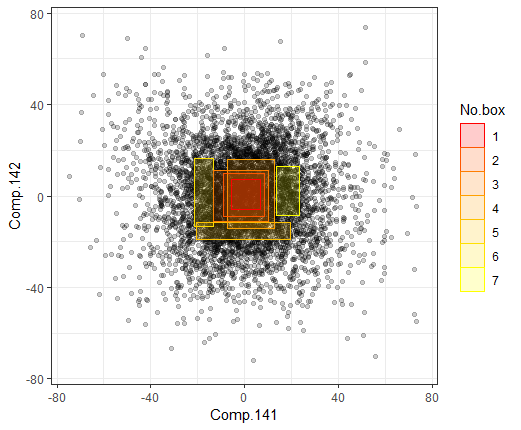}}

\subfloat[Principal  with PRIM for 9]{\includegraphics[width=4cm,height=3cm]{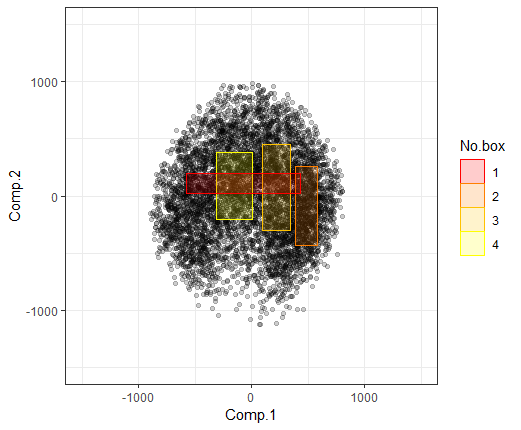}}
\subfloat[Pettiest  with PRIM for 9]{\includegraphics[width=4cm,height=3cm]{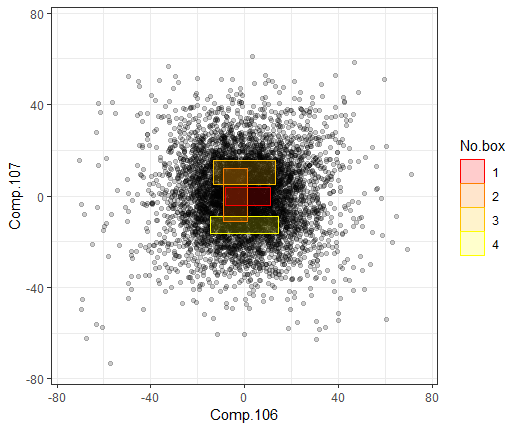}}

\caption{MNIST modeling results 5--9 (PRIM)}
\end{figure}

\begin{figure}[!t]
\centering 
\includegraphics[width=0.5\textwidth]{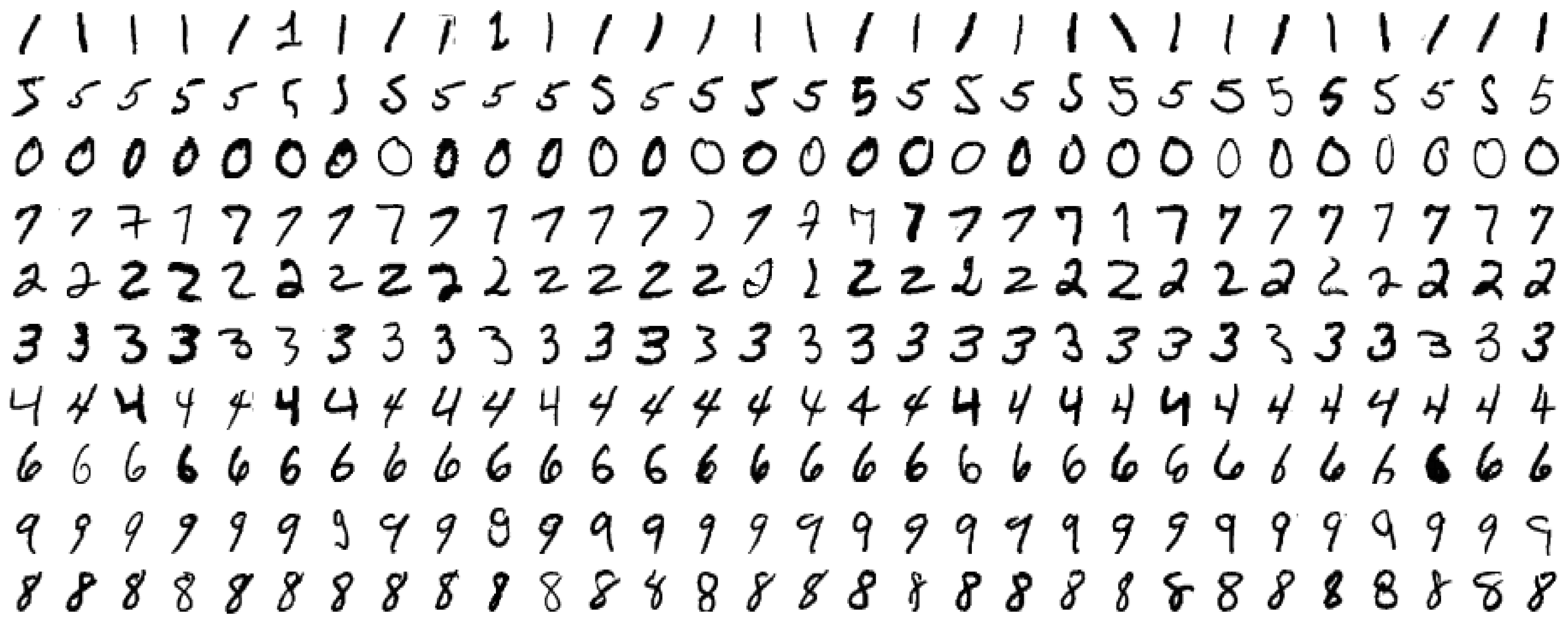}
\caption{Handwritten digits samples}
\label{Fig.digit}
\end{figure}

To see this superiority formally, we calculate the active information \eqref{actinfo} for each of the 10 digits considering the 4 different methods (PRIM with principal components, PRIM with pettiest components, fastPRIM with principal components, and fastPRIM with pettiest components), setting the base of the logarithm at 2. After, we rank the digits from highest to lowest in terms of the active information of the final region obtained. The results are shown in Table  \ref{table.AI} (we recommend to look simultaneously at Fig.\,\ref{Fig.digit}, a sample of 30 observations from the training set, to better understand this discussion).

\begin{table*}[!t]
\scriptsize
\caption{Active information by method}\label{table.AI}%{
%\resizebox{\textwidth}{!}{
\centering
  \begin{tabular}{cccccccc}
     Num.& PRIM-Principal & Num. & fastPRIM-Principal & Num. & PRIM-Pettiest  & Num. & fastPRIM-Pettiest\\
    2 & 1.90 & 2 & 1.83 & 1 & 8.08 & 1 & 7.96\\
    4 & 1.81 & 4 & 1.79 & 5 & 6.30 & 5 & 6.15\\
    1 & 1.79 & 7 & 1.36 & 0 & 6.26 & 0 & 5.97\\
    7 & 1.58 & 8 & 1.32 & 7 & 5.71 & 7 & 5.71\\
    5 & 1.47 & 3 & 1.24 & 2 & 5.44 & 2 & 5.42\\
    3 & 1.36 & 9 & 1.07 & 3 & 4.96 & 3 & 4.94\\
    8 & 1.31 & 5 & 0.89 & 4 & 4.69 & 4 & 4.70\\
    0 & 1.29 & 6 & 0.81 & 6 & 4.54 & 6 & 4.54\\
    9 & 1.24 & 0 & 0.60 & 9 & 4.06 & 9 & 4.13\\
    6 & 1.05 & 1 & -1.21 & 8 & 3.39 & 8 & 3.36
  \end{tabular}%}}
\end{table*}

According to \eqref{actinfoComp}, the $\beta$-mode is the region of the space with probability $\beta$ having the highest active information. It means that this $\beta$-mode corresponds to the region of the space with minimal hyper-volume having probability $\beta$. Taking also into account that active information was first introduced to measure how much information a programmer is adding to an algorithm in order for such algorithm to reach a target $T$ with respect to a blind search  \cite{DembskiMarks2009a, DembskiMarks2009b}, we can offer an interpretation of the findings in Table \ref{table.AI}.

The target $T$ in the particular context of our example is the handwritten pattern of each digit, say the Platonic idealization of Arabic numerals. Since such pattern for 1 is the easiest to follow (a vertical segment), it is not surprising that the active information of the mode representing 1 is about 8. In fact, there is a difference of almost two bits of information in the ranking of the first two digits. Accordingly, it is reasonable to expect the digit 0 to rank high, since its very well defined geometric representation makes for an easy pattern to follow (a circumference); its active information is close to 6. Somewhat surprisingly the digit 5 sits second, between the digits 1 and 0. Unsurprisingly, all the last five digits (3, 4, 6, 9, 8) have active information below 5, which is easily explainable in terms of the more complex shapes they have, since all, except the digit 4, involve some circular shape plus some additional pattern. As for 4, its ranking is possibly explained by the two ways there is to represent it. Notice also that the digits 6, 9, and 8, having obtained the highest estimation of $\beta$ through the cross-validation, rank in the last positions. The digit 8 ranks last, being the only digit with active information around 3 bits, and a difference of 1.6 bits with respect to the digit 9, the second to last. However, it is important to notice that even with 8 we see an active information bigger than 3 for the mode of the pettiest components; this says that the mode is at least 8 times more probable than the uniform probability of the same box. Thus, even with 8 a very well defined pattern is followed; more so with all the other digits.

Observe that the ordering of several digits in Table \ref{table.AI} is heavily twisted when we see them first in the principal components and then in the pettiest components (both with PRIM and fastPRIM). For instance, 1 ranks first with fastPRIM pettiest components but last with fastPRIM principal components, obtaining a negative active information. Also, 4 ranks seventh with fastPRIM pettiest components but second with fastPRIM principal (5 does the opposite). Nonetheless, active information will clearly show that it will be mistaken to look for the $\beta$-mode of the projection in the principal components instead of the pettiest components. To see this, notice from Table  \ref{table.AI} that we obtain an important active information gain when jumping from principal to pettiest components. Since $\beta$ is fixed for each digit, and the active information of the final region $R$ can be written as 
\begin{align*}
	\textbf I_+(R) &= \log \beta - \log \textbf U(R)\\
				&= \log \beta + \log (\text{Vol}(S)) - \log(\text{Vol}(R)),
\end{align*}
where $S$ is the sample space, the only term that is changing in determining the active information between PRIM and fastPRIM is the last term at the RHS of the previous equation, $\log(\text{Vol}(R))$. Thus, for instance, for the digit 1 evaluated with fastPRIM  the final region obtained with pettiest components is more than $2^9$ times smaller than the region found with fastPRIM principal components. Even for the digit 8, whose active information difference between fastPRIM pettiest and fastPRIM principal is minimal, we obtain that the final region with pettiest components is $\sim 2^4$ smaller than the region with principal components. The volume reduction is extremely significant. In fact, observe that many of the distributions around principal components look uniform (see, e.g., \ref{Fig.example1}(a) and (g) for the digits 1 and 3 with principal components, with the notorious exception of the digit 1), whereas the distribution of digits with pettiest components have in general more structure that gives them starry shapes.

Notice also the interesting fact that when looking at the pettiest components, PRIM does better than fastPRIM with the digits 0, 1, and 5. This can be puzzling at the beginning. Comparing fastPRIM and PRIM with principal components for the digit 1, the most extreme case, is enlightening here. Notice from Figs. \ref{Fig.example1}(c) and \ref{Fig.example2}(c), that the projected distribution in the dimension of the two principal components is not symmetric. Since fastPRIM was developed for symmetric unimodal distributions, it is not surprising that PRIM does a better job in this case. Nonetheless, it is important to notice that there is no violation of the results obtained in our theorems, since the conditions are not satisfied. Nonetheless, it is important to highlight that in spite of the data not being normal or Laplace, PRIM with pettiest components is doing only slightly better than fastPRIM, illustrating that even in these cases little to nothing is lost if we consider fastPRIM instead of PRIM. 

\begin{figure}[!t]
\centering
\subfloat[Reconstruction with principal]{\includegraphics[width=6cm,height=1cm]{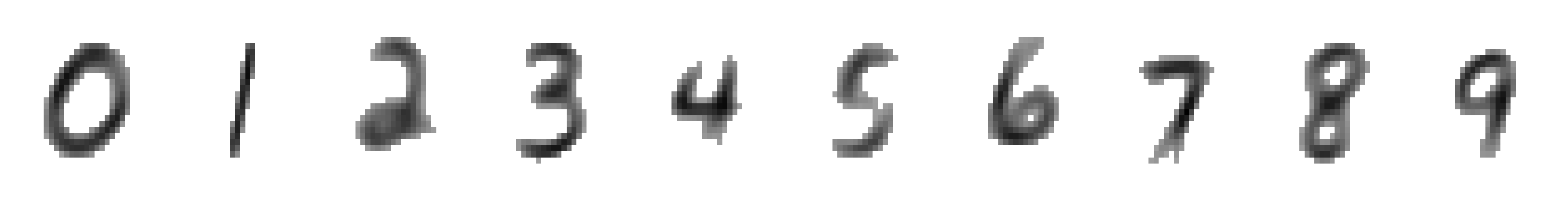}}\\
\subfloat[Reconstruction with pettiest]{\includegraphics[width=6cm,height=1cm]{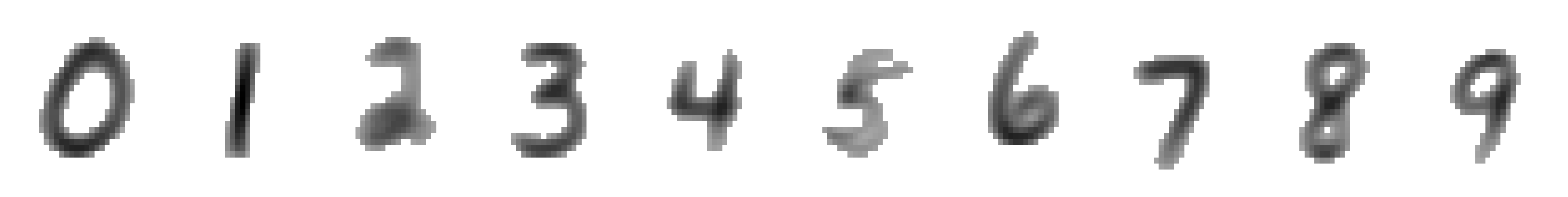}}\\
\subfloat[Averages]{\includegraphics[width=6cm,height=1cm]{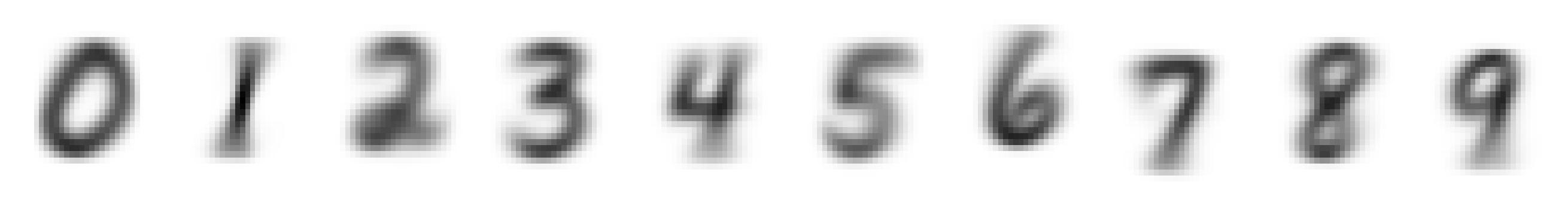}}
\caption{Reconstructed digits with fastPRIM.}\label{recons}
\end{figure}

\begin{figure*}[!t]
\centering

\subfloat[Grid of 1 with principal]{\includegraphics[width=6cm,height=5cm]{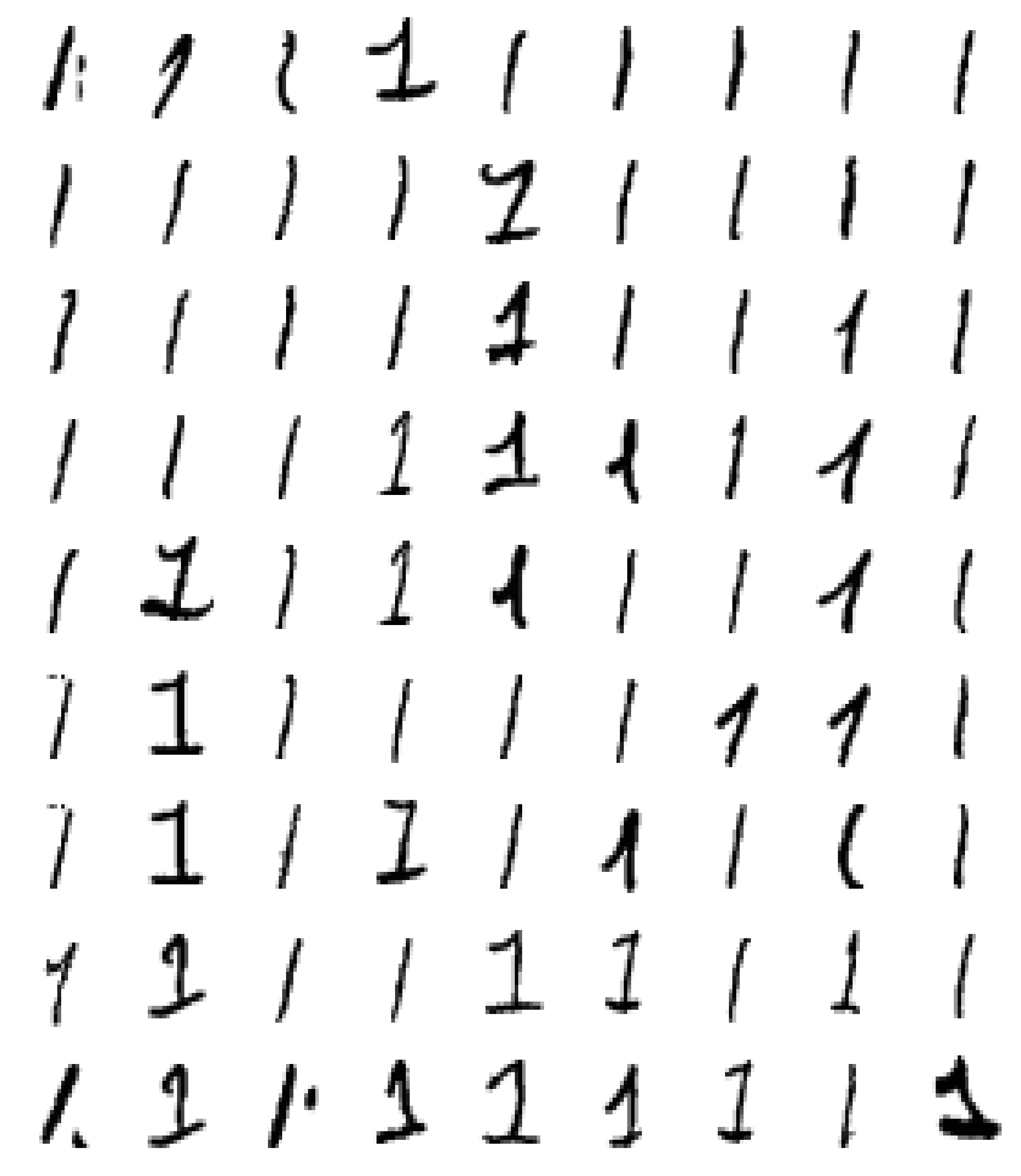}} 
\subfloat[Grid of 0 with principal]{\includegraphics[width=6cm,height=5cm]{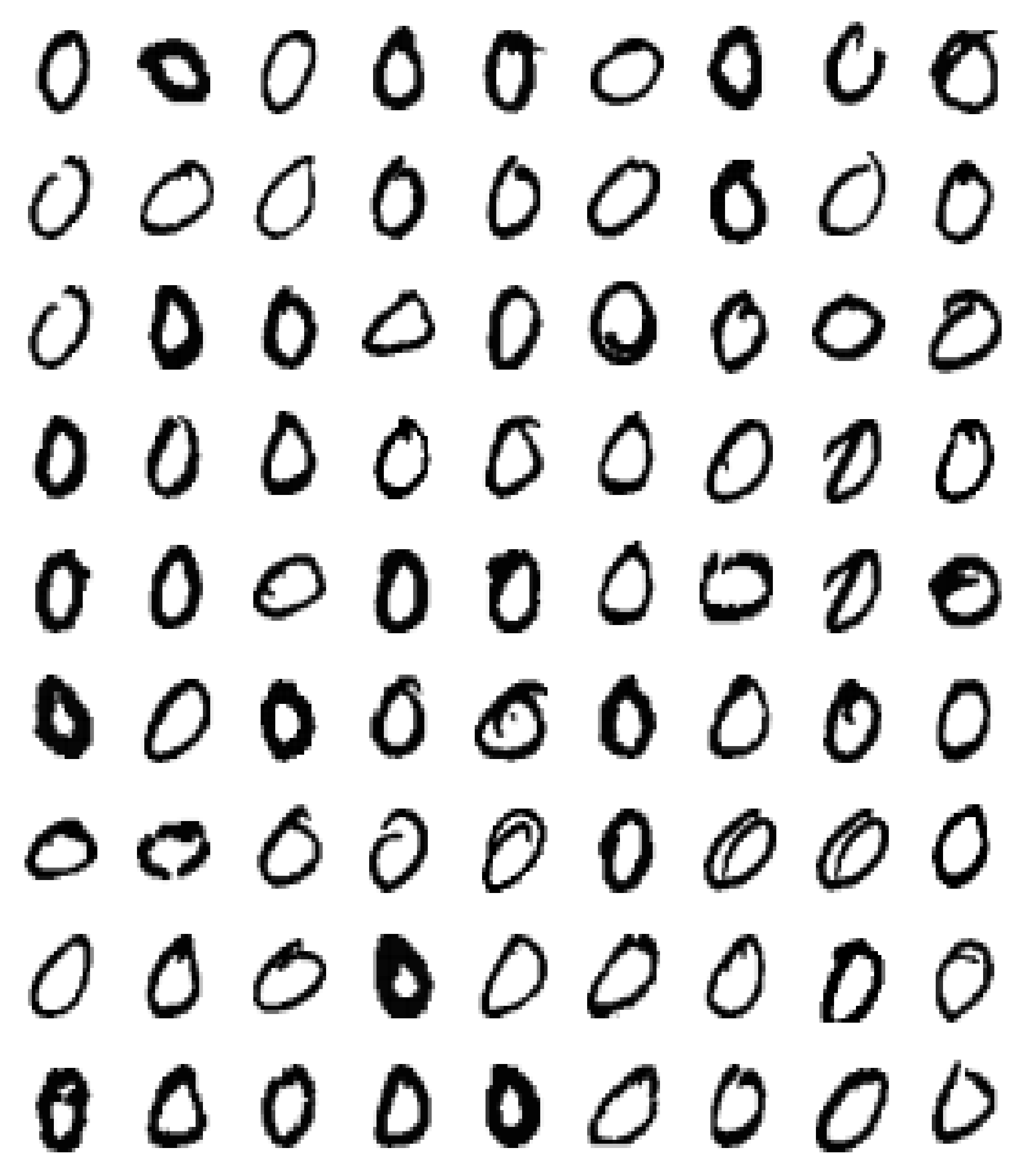}} 

\subfloat[Grid of 1 with pettiest]{\includegraphics[width=6cm,height=5cm]{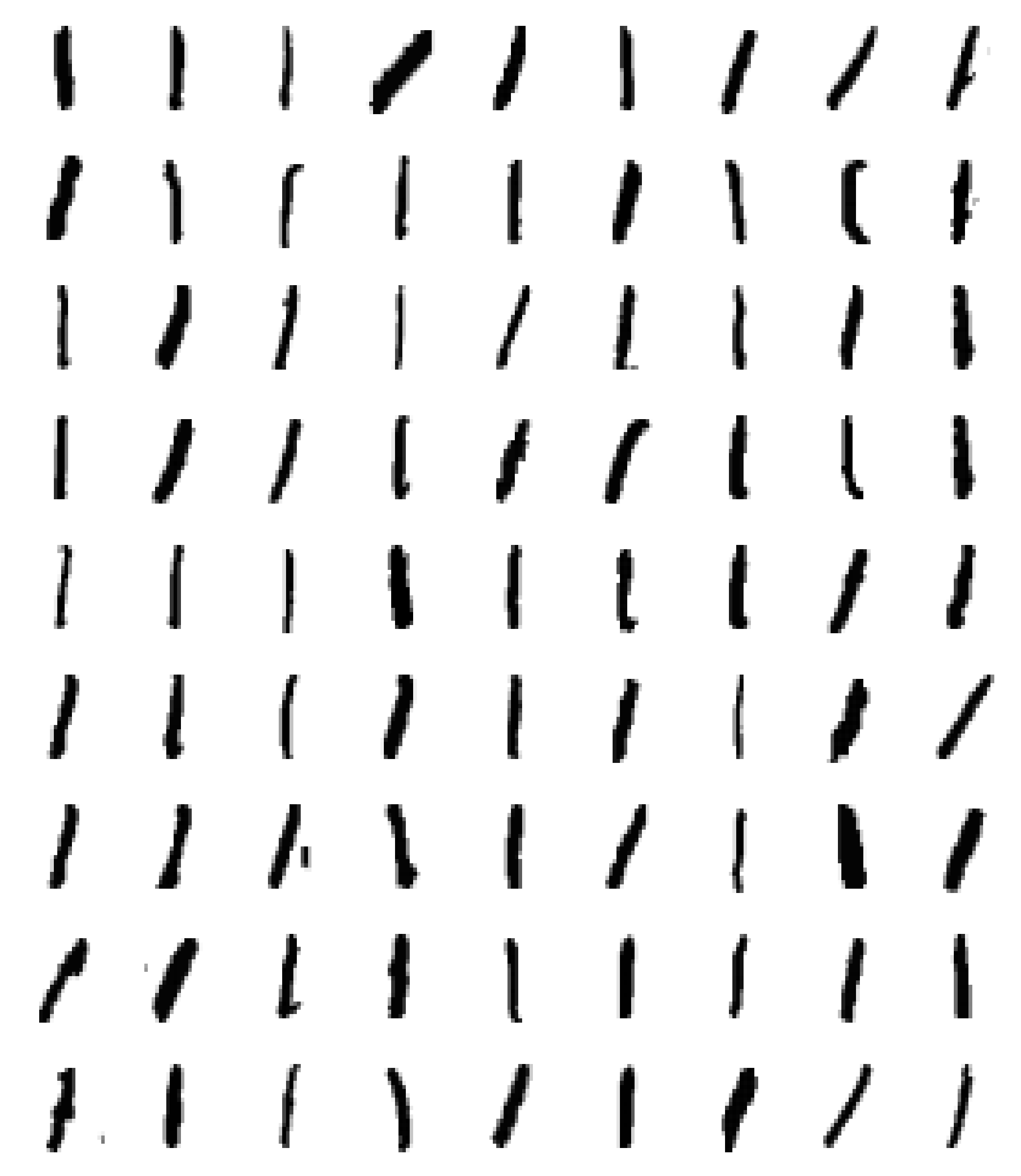}} 
\subfloat[Grid of 0 with pettiest]{\includegraphics[width=6cm,height=5cm]{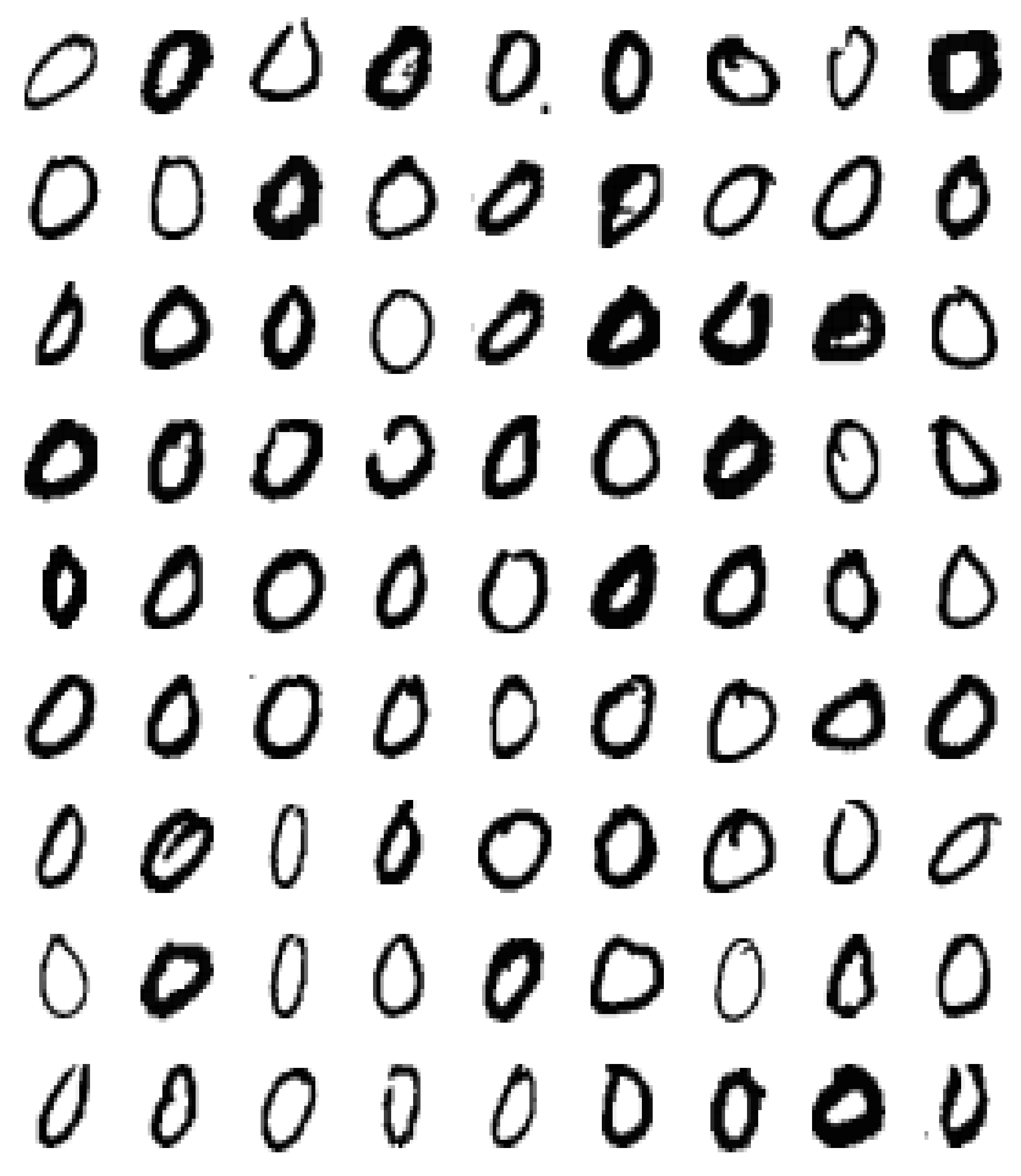}} 

\subfloat[Grid of 5 with principal]{\includegraphics[width=6cm,height=5cm]{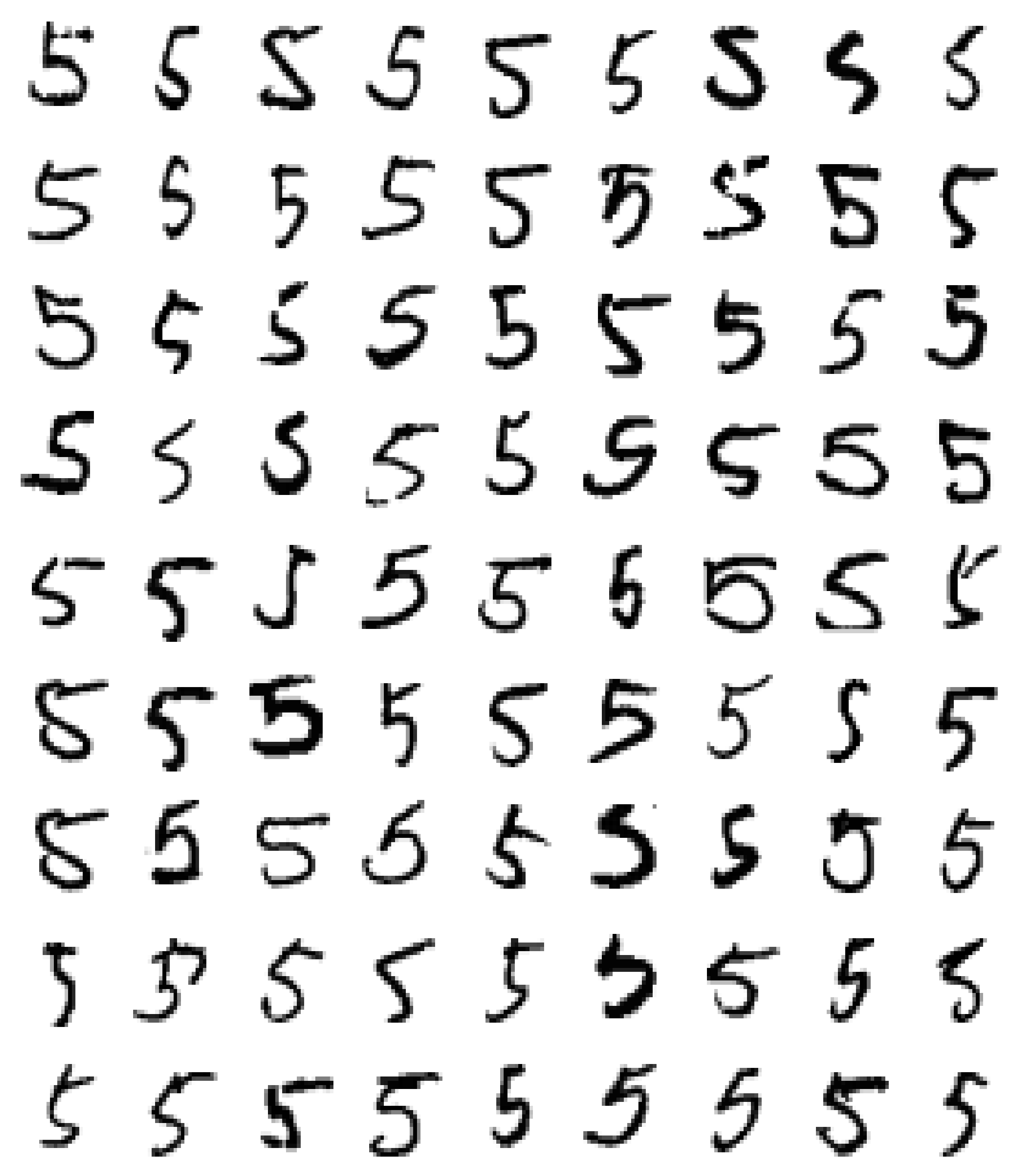}}
\subfloat[Grid of 8 with principal]{\includegraphics[width=6cm,height=5cm]{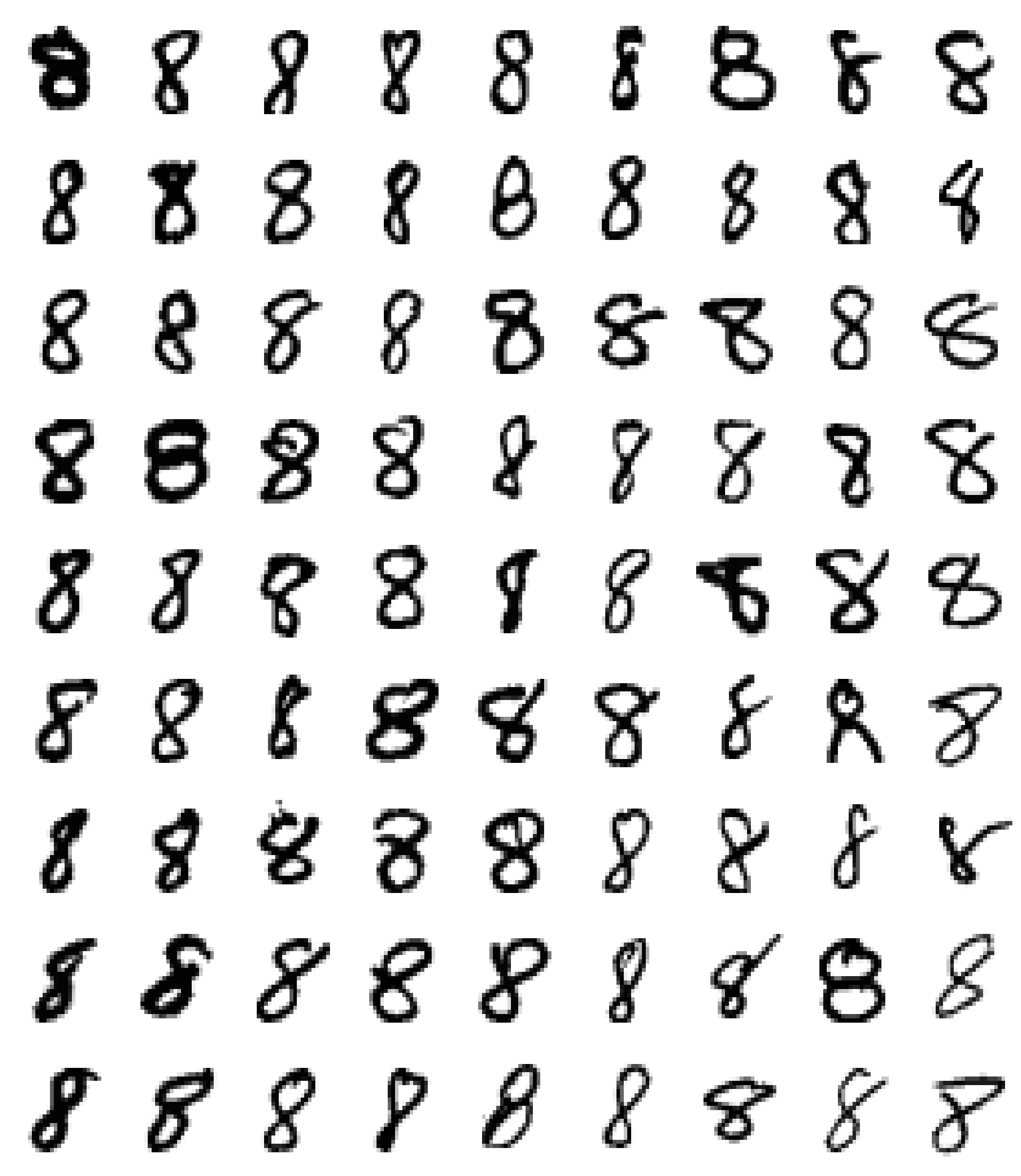}}

\subfloat[Grid of 5 with pettiest]{\includegraphics[width=6cm,height=5cm]{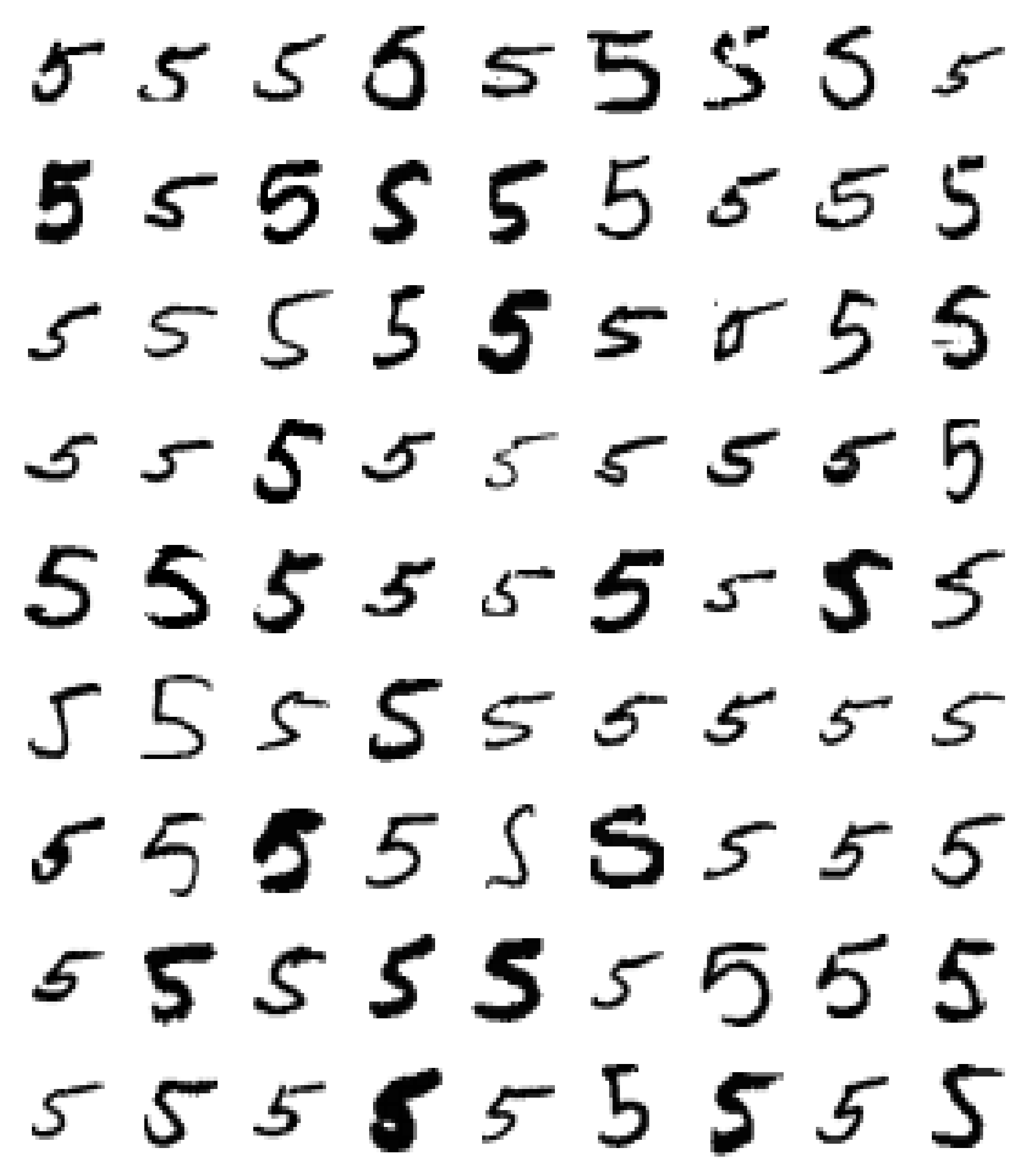}}
\subfloat[Grid of 8 with pettiest]{\includegraphics[width=6cm,height=5cm]{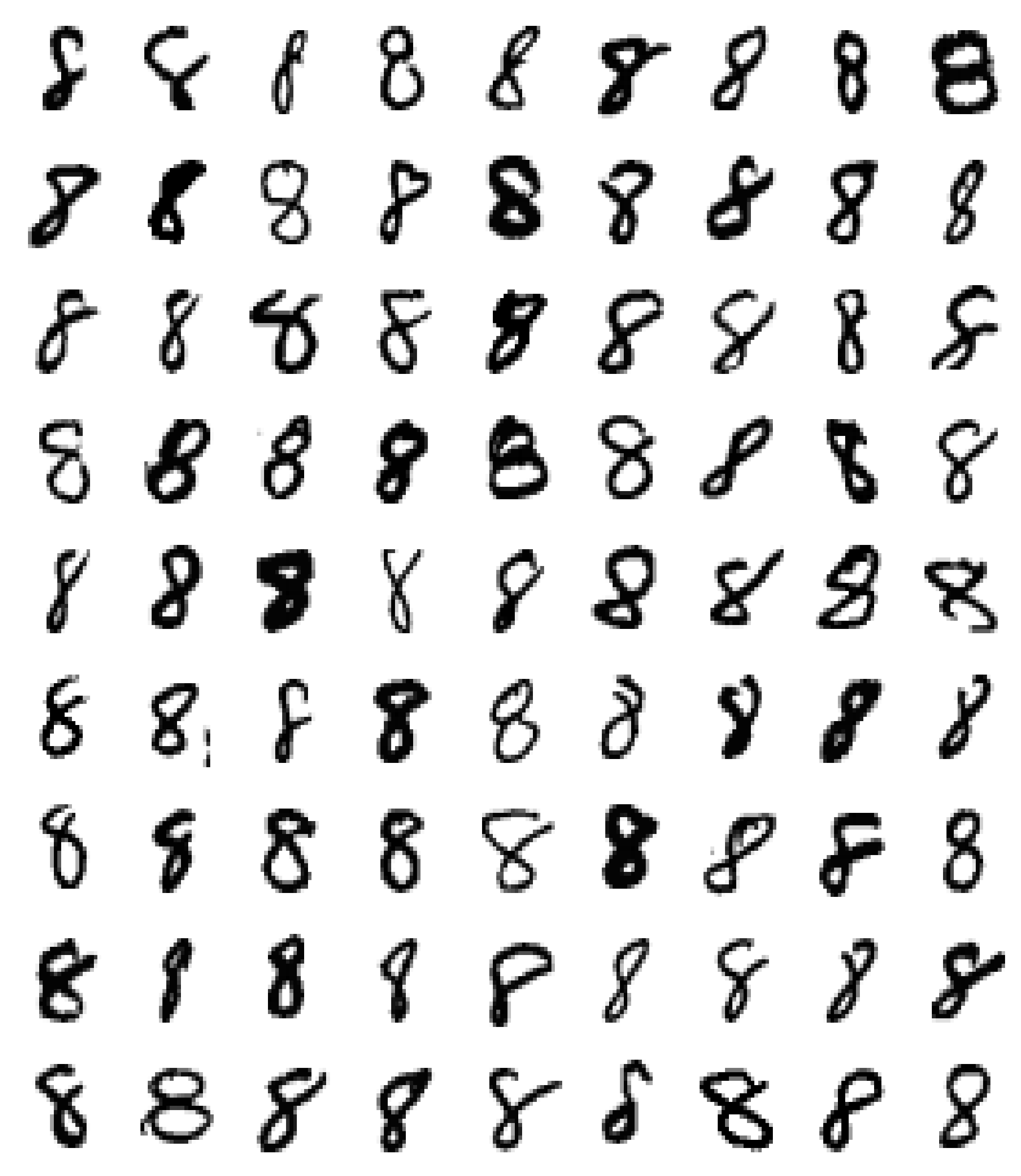}}

\caption{Equidistant points inside the $\beta$-region with fastPRIM}\label{comparison}
\end{figure*}

Finally, we propose to use pettiest components to reconstruct the number. Or, to be more accurate, we propose using the region with the highest active information in order to reconstruct the number. Our reasoning is simple. The region with the largest active information will be close to the platonic idealization of the digit (or at least a democratized version of  such platonic idealization). To see this, we show in Fig. \ref{comparison} how the digits for 1, 5, 0, and 8 look between fastPRIM with pettiest and principal components. Notice that the higher the difference in active information between the two procedures, the more obvious become to use of this strategy. For instance, for digit 1 the difference in active information between fastPRIM pettiest and principal components is above 9 bits, and it is very clear that the digits written with pettiest components are more homogeneous. The same is true for the digits 0 and 5, whose difference in active information between fastPRIM pettiest and principal is over 5 bits. Following this trend we end with 8, whose difference of active information between the two procedures is around 2 bits and visually the difference is not that obvious. Therefore, we claim that it is better to use the region with the highest active information to reconstruct the image, in this case any of the procedures considered with pettiest components. Fig. \ref{recons} illustrates this point comparing reconstruction between fastPRIM with pettiest, fastPRIM with principal, and, for control, digits created using the average: 0 and 1 seem bolder with pettiest; 2, 7, and even 9 look better finished with pettiest; 3 has a more defined round form in the lower part with pettiest, similar to what was discovered in \cite[pp.~536-539]{HastieTibshiraniFriedman2009} with principal components, but we show here that it is even better with pettiest; 5 with principal looks more like and 'S', while 5 with pettiest looks better defined; 6 seems a slightly better with principal; 8 does not seem to show too many visible differences, but when the image is zoomed in the center looks neater with pettiest. The average looks more blurred in all cases.

\begin{figure}[!t]
\centering 
\includegraphics[width=0.5\textwidth]{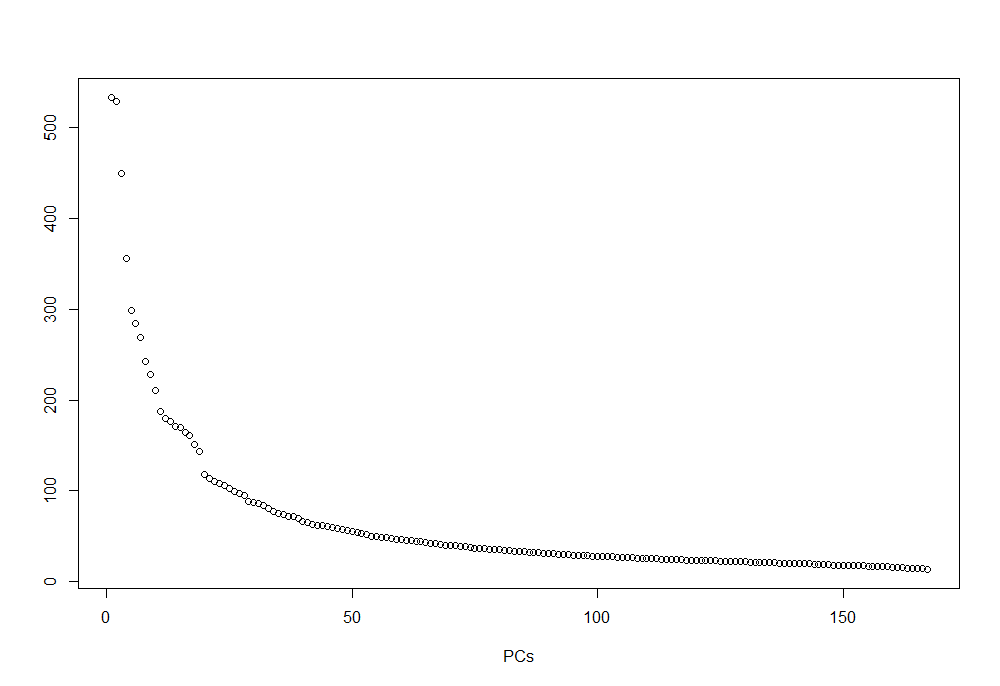}
\caption{Eigenvalues for the digit 0}
\label{Fig.EigenOrder}
\end{figure}

\begin{remark}
Spectral instability, whether small perturbations in a large matrix can lead to large fluctuations in the spectrum, is of course a point to consider. In general, the spectrum of a Hermitian matrix is stable. For non-Hermitian matrices, however, a lower bound in the least singular value is required \cite[Chapters 1.3, 2.7, and 2.8]{Tao2012}. As for numerical analyses, a QR algorithm for eigenvalues, which is numerically stable, can be used. For the particular example analyzed here, spectral instability was avoided from the beginning by the remotion of the white pixels (Fig. \ref{Fig.rank}), which not only prevented bias towards these points, but also eliminated the problem of dealing with eigenvalues in the proximity of zero. Figure \ref{Fig.EigenOrder} shows the eigenvalues for one of the digits. The last eigenvalue, corresponding to component 167 is 13.94, which is fairly large for the pettiest component. 

Moreover, for component 157 the eigenvalue is 16.99. The closeness between the last pettiest components deserves some attention. The worst scenario is going to be the case in which only the last pettiest component is selected, since small changes in the data are likely to produce changes in the ordering. However, since $p'$ components are under consideration, changes in the ordering are likely to become permutations in the ordering of such components, which is enough for mode hunting. Seen in a different way, as Fig. \ref{Fig.EigenOrder} reveals, the two leading eigenvalues are very close too, their values being 532.64 and 528.58. From a principal component analysis viewpoint, this situation is problematic if no more than the first principal component is selected. But when $p' \ge 2$ components are chosen, the problem fades away since even if small perturbations alter the order of the two leading components, both components will likely still be part of the $p'$ components selected.
\end{remark}

\section{Summary}\label{Sec:Disc}

Given the lack of robustness of the mean, even in low dimensions, and the prevalence of big data and large-dimensions analyses, mode-based statistics and learning is becoming more relevant \cite{Chacon2020}. Consequently, new theory and methods are required in order to better detect modes, but this task is difficult and elusive since kernel functions become ineffective in even not-that-high dimensions. The importance of fastPRIM in mode hunting is dictated by the central limit theorem. The setback of this approach is the curse of dimensionality, as Vapnik explains in the Introduction of his book \cite[pp.~ 4--6]{Vapnik1998}. Therefore, even though there is some optimality that is reached just by rotating the information in the direction of the eigenvalues, if mode detection methods are going to be useful they will need to reduce dimensionality. 

In this sense, even though it is well known that pettiest components can sometimes explain better a response than principal components, the latter have been treated as the ideal tool whereas the former have been considered isolated counter-examples or anomalies to be avoided. However, in this article we showed that when projecting the multivariate data in the direction of a few eigenvalues, pettiest components can be systematically used in order to find the best $\beta$-modes, provided that the data is distributed normal or Laplace. That this had not been noted before is surprising, given the centrality of the normal distribution in statistics. 

This finding goes against the general notion that principal components are more informative, since, as shown in the Introduction, the box with the smallest size also maximizes the active information relative to the uniform distribution. In fact, the theorems and the example with the MNIST dataset illustrate that principal components can totally twist the importance of the components, since active information shows that the true order is the one given by pettiest components, and this is so even when there are strong departures from normality. Using pettiest over principal components produces significant gains. 

Some questions remain open. For instance, how much can Theorems \ref{gentheo} and \ref{laplace} be extended to a more general family of distributions, involving all unimodal symmetric multivariate distributions formed by marginals corresponding to the same family?

A second extension is to consider the extent to which the marginals can depart from families of distribution (as in the counterexample in Remark 1), while our result is maintained.

A third extension in applications will be to develop numerical methods allowing the calculation of stable eigenvalues for the pettiest components. The approach taken for the MNIST example seems a good starting point, in which a pettiest components analysis was performed not for the original dataset but for an already reduced space in which decent eigenvalues were obtained even for the pettiest components. It suggests, in agreement with the theory of stability of eigenvalues, that finding a threshold above which to consider pettiest components is necessary. This procedure would be similar to the so-called ``one standard error rule'' for model selection \cite[p.~244]{HastieTibshiraniFriedman2009}, \cite[p.~214]{JamesEtAl2013}.

A fourth extension will be to develop a theory to distributions with multiple modes, which will require some partition of the space. A good starting point in this direction might be Dazard and Rao's partition in their Local Sparse Bump Hunting algorithm \cite{DazardRao2010}.

In spite of these open questions, as the real data example shows, using pettiest components instead of principal components in order to determine the best $\beta$-mode in a projected space is, in general, a wise idea.

%\section{Acknowledgements}
%
%JSR was partially supported by NSF grant DMS-1915976 and NIH grants U54 MD010722 and UL1 TR000460. 

% For fastPRIM

% for PRIM

%That being said, since the purpose of this paper was to illustrate an instance in which pettiest components are systematically more useful than principal components, we have not explored in detail other possible approaches beyond fastPRIM to correct the previous situation. Nonetheless, two generalizations are possible to get around this situation. On the one hand, Local Sparse Mode Hunting (a particular case of Local Sparse Bump Hunting \cite{DazardRao2010}) can be developed using pettiest components instead of principal components when the data is distributed as a multivariate mixture of normal distributions. On the other hand, Active Information Mode Hunting, as developed in \cite{DiazEtAl2019}, can be used in conjunction with pettiest components in order to find modes in large dimensions when the data is not normally distributed.

%\bibliographystyle{biometrika}
%\begin{thebibliography}{22}
%\expandafter\ifx\csname natexlab\endcsname\relax\def\natexlab#1{#1}\fi
%\section*{References}

\bibliographystyle{IEEEtran}

\bibliography{IEEEabrv, /Users/daniela.diaz/Documents/Research/daangapaBibliography.bib}

\begin{IEEEbiography}[{\includegraphics[width=1in,height=1.25in, clip, keepaspectratio]{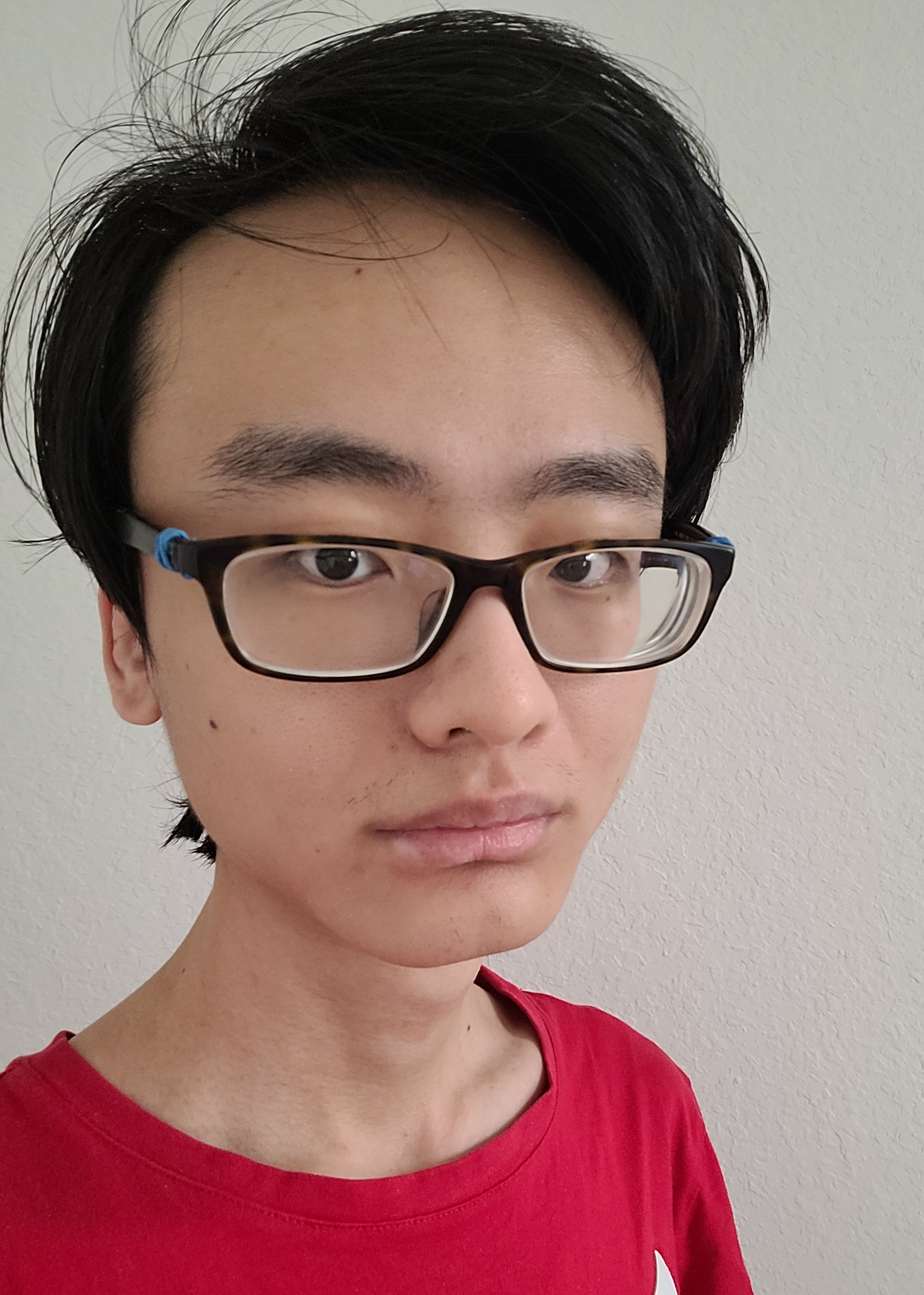}}]{Tianhao Liu} 
received his BS in Physics from Nankai University, China, in 2019, and a MS in Biostatistics from University of Miami, in 2020. He is currently working towards a PhD in Biostatistics at the University of Miami. His research interests are mainly in statistical techniques for pattern recognition and their applications to medicine.\end{IEEEbiography}

\begin{IEEEbiography}[{\includegraphics[width=1in,height=1.25in,angle =90, clip, keepaspectratio]{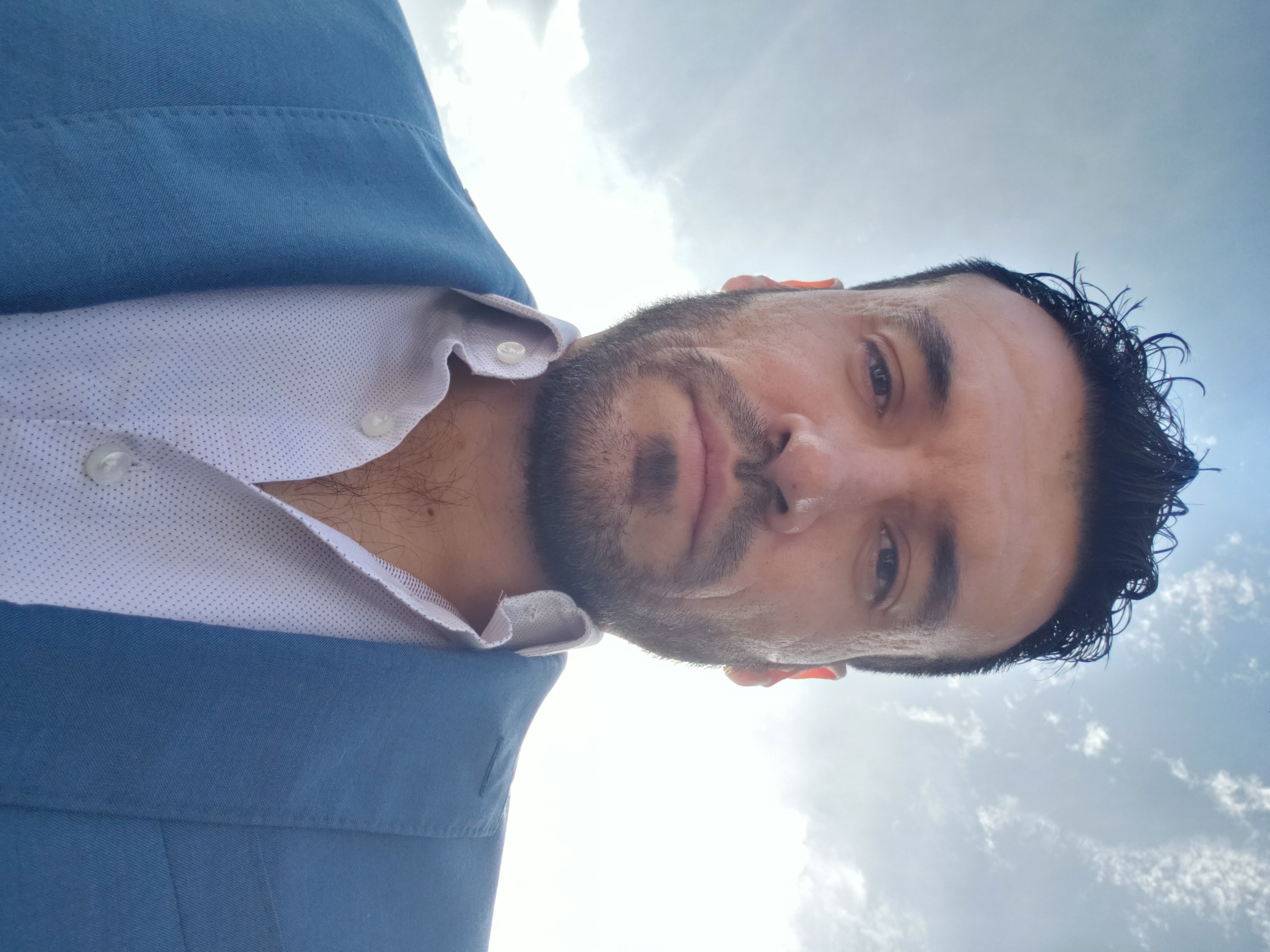}}]{Daniel Andr\'es D\'{\i}az-Pach\'on}
received his B.S. in Mathematical Statistics at \textit{Universidad Nacional de Colombia}, Colombia (2005); and his PhD in probability theory at \textit{Universidade de S\~ao Paulo}, Brazil (2009). In 2011 he moved to the University of Miami, Florida, where he was first a Postdoctoral Associate in Biostatistics (2011--2015), and then became Research Assistant Professor. His research is focused on the intersection of probability theory, statistics, machine learning, and information theory.
\end{IEEEbiography}

\begin{IEEEbiography}[{\includegraphics[width=1in,height=1in,clip, keepaspectratio]{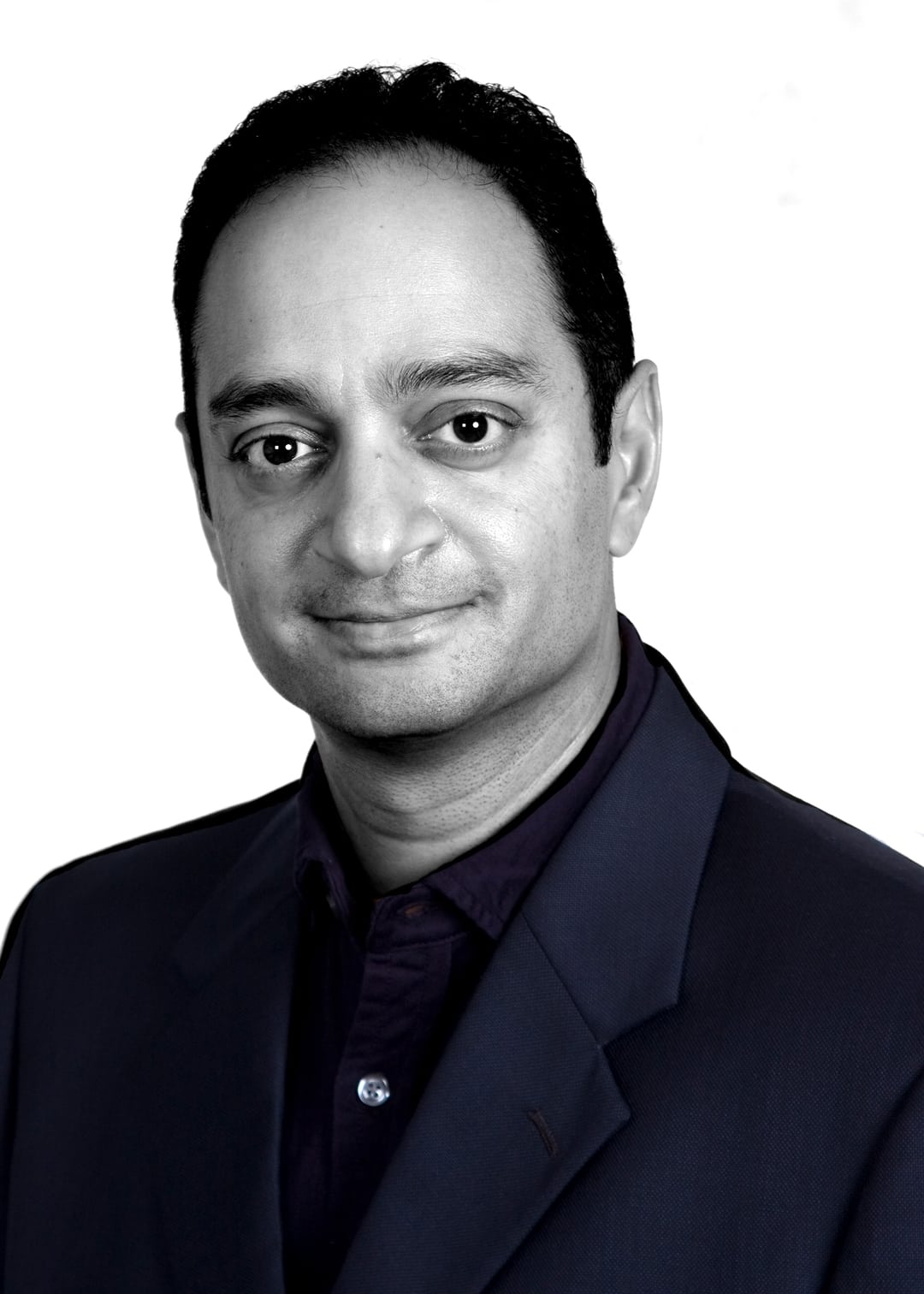}}]{J. Sunil Rao} PhD is Professor and Director of the Division of Biostatistics in the Department of Public Health Sciences, University of Miami Miller School of Medicine.  His research interests include mixed model prediction and selection, Bayesian model selection, small area estimation, machine learning and applied biostatistics.
\end{IEEEbiography}
 
\begin{IEEEbiography}[{\includegraphics[width=1in,height=1in, clip, keepaspectratio]{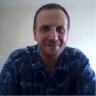}}]{Jean-Eudes Dazard} 
PhD received his PhD in Bioinformatics in 2000 from the University of Montpellier, France, following two master's degrees: in computer science in 1992 from ESIM School of Engineering, France and in Statistics in 2009 from Case Western Reserve University, USA. His research is centered on Computational/Statistical Biology. Recent focus has been in: Bump Hunting, Regularization and Variance Stabilization, Variable Selection; and Causal Regulatory Network Analysis. His interest is also in Statistical Computing and Software Development: He has authored several R packages available in CRAN and GitHub.
\end{IEEEbiography}

\end{document}